\documentclass[%
 reprint,
nofootinbib,
 amsmath,amssymb, aps,
]{revtex4-2}
   
\usepackage{graphicx}
\usepackage{dcolumn}
\usepackage{bm}
\usepackage[hidelinks,colorlinks=true,allcolors=blue]{hyperref}
\usepackage{makecell}

\usepackage{caption,subcaption}
\usepackage[x11names]{xcolor}
\usepackage{slashed}
\usepackage[utf8]{inputenc}
\usepackage[compat=1.1.0]{tikz-feynman}
\usepackage{amsmath,amssymb,multirow,graphicx}
\usepackage{amsthm}
\usepackage{braket}
\usepackage{relsize}

\newtheorem{theorem}{Theorem}

\theoremstyle{definition}

\theoremstyle{remark}

\newcommand{\calB}{\mathcal{B}}

\newcommand{\calD}{\mathcal{D}}

\newcommand{\calF}{\mathcal{F}}

\newcommand{\calO}{\mathcal{O}}

\newcommand{\calS}{\mathcal{S}}

\newcommand{\calZ}{\mathcal{Z}}
\newcommand{\e}{\mathrm{e}}
\renewcommand{\i}{\mathrm{i}}
\renewcommand{\d}{\mathrm{d}}

\newcommand{\im}{\mathop{\mathrm{im}}}

\newcommand{\Z}{\mathbb{Z}}
\newcommand{\Q}{\mathbb{Q}}
\newcommand{\R}{\mathbb{R}}
\newcommand{\C}{\mathbb{C}}
\newcommand{\Hom}{\mathrm{Hom}}

\newcommand{\Arf}{\mathrm{Arf}}

\usepackage{stackengine}

\usepackage{titlesec}  

\newcounter{subsubsubsection}[subsubsection]

\titleformat{\paragraph}[block]{\normalfont\normalsize\bfseries}{\theparagraph}{1em}{}

\titlespacing*{\paragraph}{0pt}{1.5ex plus 0.5ex minus .2ex}{0.8ex plus .2ex}

\usepackage{tikz}
\usepackage{tkz-euclide}
\usetikzlibrary{calc}
\usetikzlibrary{math}
\usetikzlibrary{decorations.markings,arrows.meta}

\usepackage{subcaption}

\begin{document}

\preprint{APS/123-QED}

\title{Bosonization versus the Nielsen-Ninomiya theorem}

\author{Saif Ullah Baig}
\email{sbaig.phys@gmail.com}
\author{Shi Chen}
\email{s.chern.phys@gmail.com}
\author{Aleksey Cherman}
\email{acherman@umn.edu}
\author{Maria Neuzil}
\email{neuzi008@umn.edu}
\affiliation{School of Physics and Astronomy, University of Minnesota, Minneapolis, MN 55455, USA}

\date{\today}

\begin{abstract}
Thanks to bosonization, bosonic lattice models can offer a lattice regularization of chiral fermions. 
We construct chiral lattice fermion operators in the 2D modified Villain scalar model and evaluate their correlation functions.  This microscopic bosonic model has an ultra-local action and an ultra-local symmetry that realizes the fermionic chiral symmetry under bosonization. 
The reconstructed lattice Dirac operator has no doublers, but is consistent with the Nielsen-Ninomiya theorem because it turns out to be non-local.  The non-locality of this derived quantity at finite lattice spacing  does not pose any obstructions to gauging the non-anomalous symmetries of the model, which is itself ultra-local.
\end{abstract}

\maketitle

\tableofcontents

\section{Introduction}

It is notoriously challenging to put massless fermions on a Euclidean
spacetime lattice while preserving chiral symmetries. This poses an
obstruction to the lattice regularization of chiral gauge theories, such as the Standard
Model. The most famous no-go theorem for lattice chiral fermions, the Nielsen-Ninomiya
theorem~\cite{Nielsen:1980rz,Nielsen:1981xu,Karsten:1980wd,Karsten:1981gd,Nielsen:1981hk,Friedan:1982nk},
roughly states that any reasonable lattice Dirac operator $\slashed{D}$ either violates
standard chiral symmetries or produces extra momentum-space zeros
corresponding to unphysical massless modes (i.e.~doublers). Standard practical
lattice fermion approaches thus either give up on preserving exact
chiral symmetries at finite lattice spacing, or tolerate unphysical
doublers. Some even do both. The modified lattice chiral symmetries that 
result from L\"uscher's reinterpretation~\cite{Luscher:1998pqa} of
the Ginsparg-Wilson relation~\cite{Ginsparg:1981bj} seem to be the best one can get with finite lattice spacing. These modified chiral symmetries are not no-site and even not ultra-local, and appear in overlap and (infinite) domain-wall
fermions discretizations~\cite{Neuberger:1997fp,Neuberger:1998wv,Hasenfratz:1998ri,Kaplan:1992bt,Shamir:1993zy}.

A different approach to lattice regularizations of chiral fermions is to abandon
the use of Grassmann fields entirely. One may first realize the desired
symmetries and 't Hooft anomalies on the lattice using bosonic fields in an
ultra-local manner,
and then hope to find a critical point where chiral fermions with the
anticipated dynamics appear. 
Such a bosonized description fundamentally eliminates the ground for doublers.
This idea relies on the same fundamental principle as symmetric mass
generation (see
e.g.~Refs.~\cite{Wang:2022ucy,You:2017ltx,You:2017mkc,Fidkowski:2009dba,
Tong:2021phe,Eichten:1985ft,Zeng:2022grc,
Golterman:2023zqf,Golterman:2025boq,Golterman:2026sox}): theories with
identical symmetry and anomaly structures are expected to be deformable
into each other. This idea has a particularly fruitful realization in
two spacetime dimensions, where explicit bosonized descriptions of
continuum fermions have long been known~\cite{Schultz:1964fv,Coleman:1974bu,Witten:1983ar} and the Coleman-Mermin-Wagner theorem renders a critical point inescapable. For
example, up to a topological manipulation, the 2D modified Villain
model~\cite{Gross:1990ub,Sulejmanpasic:2019ytl,Gorantla:2021svj}
provides a lattice regularization for a 2D massless Dirac fermion~\cite{Berkowitz:2023pnz}. 

Although several variants of this idea has been explored recently in Refs.~\cite{Berkowitz:2023pnz,Seifnashri:2023dpa,DeMarco:2023hoh,Cheng:2022sgb,Fazza:2022fss,Thorngren:2026ydw,Seifnashri:2026ema}, the finite-lattice-spacing properties of the chiral fermions realized by lattice bosonization remain to be clarified.  In this paper,
we study this question in the 2D modified Villain model. Our main
results are threefold. First, we identify lattice operators
that flow to continuum Weyl fermions and compute their exact two-point
functions. Second, by inverting these two-point functions, we
reconstruct a lattice Dirac operator and show that it is doubler-free
but non-local: it has a single Dirac zero and momentum-space poles, but no extra Dirac
zeros~\cite{Shamir:1993bi,Gurarie:2011qqc,Xu:2021ztz,Golterman:2023zqf,Lu:2023cev,Golterman:2025boq}.%
\footnote{This resembles SLAC/Stacey
fermions~\cite{Drell:1976mj,Rabin:1981nm,Weinstein:1982ht,Stacey:1981ki,Stacey:1983cb,Stacey:1983me,Stacey:1985rqq} at the level of the induced
Dirac operator, but here the non-local object is a reconstructed inverse
propagator rather than a microscopic fermion action. The underlying
bosonic lattice model is ultra-local.}
Third, we show that the lattice Weyl operators have non-vanishing
connected four-point functions at finite lattice spacing, although the
corresponding interaction is irrelevant in the continuum limit. These
results expose the way in which the bosonized model evades the Nielsen-Ninomiya
theorem: it gives up a local Dirac kernel.  But it does not give up the ultra-locality of the
microscopic bosonic theory. 

Readers familiar with 
the Nielsen-Ninomiya theorem on Euclidean spacetime lattice and the subtle global aspects of 2D
boson-fermion dualities may skip the review sections and  jump directly to
Section~\ref{sec:modified_Villain_section}.
We exclusively work with the Euclidean signature throughout.

\section{Review of the Nielsen-Ninomiya Theorem}
\label{sec:Nielsen-Ninomiya}

\begin{table*}[t]
\footnotesize
\setlength{\tabcolsep}{4pt}
\renewcommand{\arraystretch}{1.15}
\begin{ruledtabular}
\begin{tabular}{lcccccl}
Discretization & (A)  & (B) & (C) & (D) & Doublers? &
\makecell[l]{Comments} \\ 
\hline
naive fermions & yes & yes & yes & yes & yes &
\makecell[l]{$2^n - 1$ doublers in $n$D} \\
\hline
staggered fermions~\cite{Susskind:1976jm,Bock:1992yr,Catterall:2023nww} & no & no & yes & yes & yes &
\makecell[l]{doublers are reduced, but tastes remain} \\
\hline
Wilson fermions~\cite{Wilson1977} & yes & no & yes & yes & no &
\makecell[l]{Wilson term breaks the standard chiral structure} \\ \hline
\makecell[l]{overlap/infinite
domain wall\\fermions
\cite{Ginsparg:1981bj,Kaplan:1992bt,Shamir:1993zy,Neuberger:1997fp,Luscher:1998pqa}} & yes & no &
yes & yes & no & \makecell[l]{modified chiral symmetry from the Ginsparg-Wilson\\relation} \\ \hline
\makecell[l]{SLAC/Stacey fermions~\cite{Drell:1976mj,Rabin:1981nm,Weinstein:1982ht,Stacey:1981ki,Stacey:1983cb,Stacey:1983me,Stacey:1985rqq}} &
yes & yes & no & yes & no & \makecell[l]{non-local lattice model from a
non-local Dirac operator} \\ \hline
\makecell[l]{bosonization~\cite{Berkowitz:2023pnz,Seifnashri:2023dpa,DeMarco:2023hoh,Cheng:2022sgb,Fazza:2022fss,Seifnashri:2026ema} (this work)} & yes & yes & no & yes & no
& \makecell[l]{ultra-local lattice model with a non-local reconstructed \\
Dirac operator}
\end{tabular}
\end{ruledtabular}
\caption{\label{tab:NN_scorecard}A Nielsen-Ninomiya scorecard for
several lattice fermion constructions, in terms of the assumptions of the theorem reviewed in the main text of Section~\ref{sec:Nielsen-Ninomiya}: (A) translation invariance, (B) chirality and hermitianity $\slashed{D}=\gamma^\mu D_{\mu}$, (C) locality, and (D) correct small-$p$ continuum limit $\widetilde{D}_{\mu}(p)\sim p_{\mu}$.}
\end{table*}

We review the Euclidean-spacetime-lattice version of the Nielsen-Ninomiya theorem~\cite{Karsten:1980wd,Karsten:1981gd}.
It is a direct corollary of the Poincar\'e-Hopf theorem, which
characterizes an obstruction to everywhere nonzero continuous vector
fields. Let us consider an $n$D closed oriented manifold $M$ and a
rank-$n$ real vector bundle $V_M$ on $M$. If a section $X$ of $V_M$ is
continuous and nonzero everywhere except at isolated points
$\{p_i\}\subset M$, the Poincar\'e-Hopf theorem states that (see
e.g.~Ref.~\cite{MilnorDifferentiableViewpoint})
\begin{equation}
    \chi(V_M) = \sum_{i} \mathrm{ind}_{p_i}X\,.
\end{equation}
The left-hand side is the Euler characteristic of $V_M$, a topological
invariant. The right-hand side sums the index of $X$ at each $p_i$,
defined as the winding number of $X$ on an $S^{n-1}$ around $p_i$. For
example, if we take $V_M$ to be the tangent bundle $TM$ of $M$, the
Poincar\'e-Hopf theorem famously implies that you cannot smoothly comb a
hairy ball, since $\chi(TS^2)=2$, but you can smoothly comb a hairy
doughnut, since $\chi(TT^2)=0$.

An infinite $n$D lattice $\Z^n$ is naturally associated with an $n$D
closed manifold, namely its Fourier transform 
\begin{equation}
    T^n \simeq \left(\frac{\R}{2\pi\Z}\right)^n.
\end{equation}
We do not include the lattice spacing $a$ in the Fourier transform, so
elements in $T^n$ are dimensionless momenta. Now let us consider a
dimensionless lattice Dirac kernel $\slashed{D}(x,y)$ with
$x,y\in\Z^n$ such that:
\begin{itemize}
    \item[\textbf{(A)}] $\slashed{D}(x,y)$ is invariant under lattice
    translation.
    \item[\textbf{(B)}] $\slashed{D}(x,y)=\gamma^{\mu}D_{\mu}(x,y)$ for
    some hermitian $D_{\mu}(x,y)$, namely $D_{\mu}(x,y)=D_{\mu}^*(y,x)$. 
    \item[\textbf{(C)}] $\slashed{D}(x,y)$ is local, meaning
    $\slashed{D}(x,y)$ with $|x-y|>R$ is bounded by $C\e^{-M|x-y|}$ for
    some $R,C,M>0$.
    \item[\textbf{(D)}] $\slashed{D}(x,y)/a^{n+1}$ reduces to
    $i\slashed{\partial}\delta^{(n)}(x\!-\!y)$ in the continuum limit
    $a\to0$.
\end{itemize}
First, translation invariance (A) implies
\begin{equation}
    \slashed{D}(x,y) = \int_{T^n} \frac{\d^n p}{(2\pi)^n}\,\e^{ip\cdot(x-y)} \widetilde{\slashed{D}}(p)\,.
\end{equation}
Second, the assumption (B) implies that $\widetilde{\slashed{D}}=\gamma^{\mu}\widetilde{D}_{\mu}$ and $\widetilde{D}$ is a section of $\R^n(T^n)$, the trivial rank-$n$ real vector bundle on $T^n$. 
Third, the locality condition (C) implies
that $\widetilde{D}$ is analytic, and thus continuous, everywhere on
$T^n$. Fourth, the continuum limit (D) requires
\begin{equation}
    \widetilde{D}_{\mu}(p) \sim p_{\mu} + \calO(|p|^2)\,,
\end{equation}
meaning that $\widetilde{D}$ has a zero at $p=0$ with index $+1$
corresponding to the desired chiral fermion. 

The Poincar\'e-Hopf theorem then implies that $\widetilde{D}$ has at
least another zero somewhere with $p\neq0$, such that the sum of the
indices of all zeros equals $\chi(\R^n(T^n))=0$. For any $p\in T^n$,
$\widetilde{D}(p)=0$ if and only if $\widetilde{\slashed{D}}(p)$ has a
zero eigenvalue, because the gamma-matrix algebra implies
\begin{equation}
    \det\widetilde{\slashed{D}}(p) = \left|\widetilde{D}(p)\right|^{2^{\lfloor n/2\rfloor}}.
\end{equation}
Hence an extra zero of $\widetilde{D}$ corresponds to an extra zero
eigenvalue of $\widetilde{\slashed{D}}$, which means an extra massless
fermion, usually called a \emph{doubler}. We thus obtain the basic
version of the Nielsen-Ninomiya theorem on a Euclidean lattice of
arbitrary spacetime dimension:
\begin{theorem}
    On an infinite $n$D lattice $\Z^n$, a lattice Dirac operator
    $\slashed{D}$ that satisfies the conditions (A), (B),
    (C), and (D) necessarily has a doubler.
\end{theorem}
One can sometimes obtain refined versions of this theorem by replacing
$\gamma^{\mu}$ with other matrices. For instance, one can consider a 4D
Weyl version of the Nielsen-Ninomiya theorem using
$\bar{\sigma}^{\mu}$. We focus on the basic
version of the Nielsen-Ninomiya theorem reviewed above in this paper.

To get rid of the doubler(s) while preserving the fundamental properties
(A), (C), and (D), one can violate (B) by contaminating $\slashed{D}$
with other matrices linearly independent of $\gamma^{\mu}$. Then
$\widetilde{D}$ has rank $>n$ and the Poincar\'e-Hopf theorem ceases to
apply. This inevitably destroys all on-site chiral symmetries ---
defined as any on-site symmetries that forbid fermion mass terms --- and
their 't Hooft anomalies.
Table~\ref{tab:NN_scorecard} summarizes how
several common lattice fermion constructions relate to the assumptions
above.

If the modified Dirac operator satisfies the Ginsparg-Wilson
relation~\cite{Ginsparg:1981bj}, one can find certain non-ultra-local versions of lattice chiral symmetries and encode their
anomalies in the lattice path integral
measure~\cite{Neuberger:1997fp,Neuberger:1998wv,Hasenfratz:1998ri,
Luscher:1998pqa,Luscher:1998du,Luscher:1999un}; see
Refs.~\cite{Chatterjee:2024gje,Clancy:2023ino,Singh:2025wet,Singh:2025sye}
for recent discussions. 

\section{Review of Coleman's 2D boson-fermion duality}
\label{sec:Coleman}

One should not view a boson-fermion duality as a literal equivalence
between a bosonic theory and a fermionic theory. They cannot be
equivalent due to fundamental differences: a fermionic theory
has local operators of half-integer spins and its partition function
depends on a choice of spin structure on spacetime, while a bosonic
theory only has local operators with integer spins and can live on
non-spin spacetimes.

Instead, a boson-fermion duality relates two different path-integral
descriptions of the same theory, one written using bosonic fields and
the other written using fermionic fields. A path integral over bosonic
fields can define a fermionic theory if it incorporates appropriate
topological terms that are sensitive to the spin structure. Conversely,
a path integral over fermionic fields has a chance to define a bosonic
theory if the fermionic parity $(-1)^F$ is dynamically gauged.

People have found various kinds of boson-fermion dualities on 2D
spacetime. Here we review a celebrated 2D boson-fermion duality
discovered by Coleman~\cite{Coleman:1974bu}, but from a modern point of view; c.f.~Refs.~\cite{Tachikawa:2018cer,Thorngren:2018bhj,Karch:2019lnn,Ji:2019ugf}. We shall interpret the duality as an equivalence between two
different path-integral descriptions of the same \textit{fermionic}
theory. The bosonic description of this fermionic theory is a 2D compact
boson with a ``spin $\theta$-angle,'' while the fermionic description is
a 2D massless Dirac fermion.

\subsection{Bosonic description}
\label{sec:bosonic_side}

Let us start with a 2D bosonic theory defined by the path integral over
a $2\pi$-periodic scalar field $\phi$:
\begin{equation}\label{eq:compact_scalar}
    \int\calD\phi\,\exp\left\{ -\frac{R^2}{4\pi}\int\d^2x\, |\partial_{\mu}\phi|^2\right\}.
\end{equation}
This is a conformal field theory with central charges $c=\bar{c}=1$.
It has a shift symmetry $U(1)_{S}$ that shifts
$\phi\to\phi+\alpha$ and a winding symmetry $U(1)_{W}$ due to
$\pi_1(S^1)=\Z$. They have a mixed 't Hooft anomaly characterized by the
3D bosonic invertible theory
\begin{equation}\label{eq:anomaly_boson}
    \exp\left\{\frac{i}{2\pi}\int A_W\wedge\d A_S\right\}
\end{equation}
for background $U(1)_{W}\times U(1)_{S}$ gauge field $A_W, A_S$ on 3D
oriented manifolds. T-duality maps $R\to1/R$ and exchanges
$U(1)_{W}\leftrightarrow U(1)_{S}$. Hence $R=1$ is self-dual, and
T-duality yields a $\Z_2$ symmetry there.

\subsubsection{Spin \texorpdfstring{$\theta$}{theta}-angle}

The topology of an $S^1$ scalar brings not only the $U(1)_{W}$ symmetry
but also topological $\theta$-angles. On 2D oriented and spin spacetime
manifolds, the $\theta$-angles are classified by the bordism groups (see
e.g.~Refs.~\cite{Freed:2017rlk,Lee:2020ojw,Chen:2022cyw,Chen:2023czk})
\begin{subequations}
\begin{gather}
    \Hom\Bigl(\widetilde{\Omega}_2^{SO}(S^1),U(1)\Bigr)=0\,,\label{eq:no_bos_theta}\\
    \Hom\Bigl(\widetilde{\Omega}_2^{Spin}(S^1),U(1)\Bigr)=\Z_2\,,\label{eq:yes_ferm_theta}
\end{gather}
\end{subequations}
respectively. All 2D orientable manifolds are spinnable, but in general
they allow multiple spin structures. As suggested by
$\d\phi\wedge\d\phi=0$, Eq.~\eqref{eq:no_bos_theta} implies that there
are no $\theta$ angles that depend only on the orientation of spacetime.
However, Eq.~\eqref{eq:yes_ferm_theta} implies that an $S^1$ scalar does
allow a nontrivial $\theta$-angle which is sensitive to the choice of
a spin structure.

This nontrivial spin $\theta$-angle can be expressed in terms of the
$\Z_2$-valued Arf invariant.%
\footnote{See e.g.
Refs.~\cite{Tachikawa:2018cer,Thorngren:2018bhj,Karch:2019lnn,Ji:2019ugf} for
useful reviews about the Arf invariant.}
For a spin structure $s$ on a closed 2D oriented manifold, we have
$\Arf(s)=0$ if the closed 2D spin manifold can bound a compact 3D spin
manifold; otherwise we have $\Arf(s)=1$. For example, on a torus $T^2$, we have
$\Arf(AA)=\Arf(AP)=\Arf(PA)=0$ and $\Arf(PP)=1$, where $A$ and $P$ stand
for the anti-periodic and periodic spinor boundary conditions,
respectively, along the two $S^1$ factors of $T^2$.

The nontrivial spin $\theta$-angle is expressed in terms of the Arf
invariant as
\begin{equation}
    \Theta_s(\phi) \equiv (-1)^{\Arf\left(s+\left[\frac{\d\phi}{2\pi}\right]_2\right) + \Arf(s)}, \label{eq:theta_spin}
\end{equation}
where
\begin{equation}
    \left[\frac{\d\phi}{2\pi}\right]_2 \in H^1(-,\Z_2)
\end{equation}
is the mod-2 reduction of $\phi$'s deformation class
\begin{equation}
    \left[\frac{\d\phi}{2\pi}\right] \in H^1(-,\Z)\,.
\end{equation}
For example, on a torus $T^2$, $\phi$'s deformation classes are captured
by a pair of integers $(a,b)$ measuring the winding numbers of $\phi$ on
the two $S^1$ factors of $T^2$. For the deformation class $(a,b)$, the
spin $\theta$-angle~\eqref{eq:theta_spin} evaluates to
\begin{equation}\label{eq:torus_theta}
\begin{gathered}
    \Theta_{AA}(\phi) = (-1)^{ab},\quad
    \Theta_{AP}(\phi) = (-1)^{ab+a},\\ \Theta_{PA}(\phi) = (-1)^{ab+b},\ \ 
    \Theta_{PP}(\phi) = (-1)^{ab+a+b}.
\end{gathered}
\end{equation}
The path integral with the spin $\theta$-angle,
\begin{equation}\label{eq:compact_scalar_spin}
    \int\calD\phi\,\exp\left\{ -\frac{R^2}{4\pi}\int\d^2x\, |\partial_{\mu}\phi|^2\right\}\Theta_s(\phi),
\end{equation}
produces partition functions that depend on the spin structure of
spacetime, and thus yields a fermionic theory.

It is now well-known that one can apply a certain topological
manipulation to any 2D bosonic theory with a non-anomalous $\Z_2$
symmetry to convert it into a 2D fermionic
theory~\cite{Tachikawa:2018cer,Thorngren:2018bhj,Karch:2019lnn,Ji:2019ugf}. We
review this construction in Appendix~\ref{app:tetrahedron}. There we
also show that applying this topological manipulation to the bosonic
theory~\eqref{eq:compact_scalar} with respect to the symmetry
$\Z_2\subset U(1)_{W}$ is precisely equivalent to activating the spin
$\theta$-angle~\eqref{eq:theta_spin}, and thus produces the fermionic
theory~\eqref{eq:compact_scalar_spin}.

\subsubsection{Physical effect}

Since the spin
$\theta$-angle~\eqref{eq:theta_spin} is trivial on $S^2$, it does not affect local dynamics,
and only affects the global structure. Hence the fermionic
theory~\eqref{eq:compact_scalar_spin} is still a conformal field theory
with central charges $c=\bar{c}=1$.
Evaluating the path
integral~\eqref{eq:compact_scalar_spin} on a flat torus,
\begin{equation}
    T^2 = \frac{\C}{L(\Z+\tau\Z)},\qquad L>0,\,\im\tau>0,
\end{equation}
e.g.~following Ref.~\cite[Sec.~8.1]{Ginsparg:1988ui} but with the insertion of the phase factors~\eqref{eq:torus_theta}, we obtain a torus partition function
$\calZ_s(\tau)$ that is independent of the scale $L$, and only depends
on the modular parameter $\tau$ and the spin structure $s$:

\begin{subequations}\label{eq:torus_Z}
\begin{gather}
    \calZ_{AA}(\tau) = \sum_{q_w\in\Z}\sum_{q_s\in\Z+\frac{q_w}{2}} \chi_{q_w,q_s}^R\!(\tau)\,\bar{\chi}_{q_w,q_s}^R\!(\bar{\tau}), \label{eq:torus_Z_AA}\\
    \calZ_{AP}(\tau) = \sum_{q_w\in\Z}\sum_{q_s\in\Z+\frac{q_w}{2}} \!\!(-1)^{q_w}\chi_{q_w,q_s}^R\!(\tau)\,\bar{\chi}_{q_w,q_s}^R\!(\bar{\tau}), \label{eq:torus_Z_AP}\\
    \calZ_{PA}(\tau) = \sum_{q_w\in\Z}\sum_{q_s\in\Z+\frac{q_w+1}{2}} \chi_{q_w,q_s}^R\!(\tau)\,\bar{\chi}_{q_w,q_s}^R\!(\bar{\tau}), \\
    \!\calZ_{PP}(\tau) = \!\sum_{q_w\in\Z}\sum_{q_s\in\Z+\frac{q_w+1}{2}} \!\!\!\!(-1)^{q_w}\chi_{q_w,q_s}^R\!(\tau)\,\bar{\chi}_{q_w,q_s}^R\!(\bar{\tau}),
\end{gather}
\end{subequations}
where
\begin{subequations}\label{eq:character}
\begin{gather}
    \chi_{q_w,q_s}^R(\tau)\equiv \frac{q^{\frac{1}{4}\left(q_wR + q_s/R\right)^2}}{\eta(\tau)}\,,\\ 
    \bar{\chi}_{q_w,q_s}^R(\bar{\tau})\equiv \frac{\bar{q}^{\frac{1}{4}\left(q_wR - q_s/R\right)^2}}{\bar{\eta}(\bar{\tau})},
\end{gather}
\end{subequations}
with $q\equiv\e^{2\pi i\tau}$ and the Dedekind eta function
$\eta(\tau)$. If we turn on flat background  $U(1)_{W}$ and $U(1)_{S}$
gauge fields with holonomies $\e^{i\alpha}$ and $\e^{i\beta}$ along the
$\tau$ direction, respectively, Eq.~\eqref{eq:torus_Z} gets an insertion
of
\begin{equation}
    \e^{i\alpha q_w + i\beta q_s}\,.
\end{equation}
Therefore, $q_w$ and $q_s$ in Eq.~\eqref{eq:torus_Z} are actually $U(1)_{W}$
and $U(1)_{S}$ charges, respectively.

The bosonic theory~\eqref{eq:compact_scalar} has symmetry
$U(1)_{W}\times U(1)_{S}\times SO(2)_{\text{geo.}}$, where
$SO(2)_{\text{geo.}}$ stands for geometric rotation. But the fermionic
theory~\eqref{eq:compact_scalar_spin} has a different symmetry.
Equation~\eqref{eq:torus_Z_AA} implies that local operators (i.e.~the
ones in the Neveu-Schwarz sector) have $U(1)_{W},U(1)_{S}$ charges
\begin{equation}
    q_w=q_{\ell}+q_r\,,\quad q_s=\frac{q_{\ell}-q_r}{2}\,,\qquad q_{\ell},q_r\in\Z\,.
\end{equation}
Equation~\eqref{eq:torus_Z_AP} implies that odd-$q_w$ local operators
are fermionic while even-$q_w$ local operators are bosonic. As a result, the symmetry
of the fermionic theory~\eqref{eq:compact_scalar_spin} has to be a
nontrivial extension of $U(1)_{S}\times SO(2)_{\text{geo.}}$ by
$U(1)_{W}$, given by
\begin{equation}\label{eq:fermionic_symmetry}
    \frac{U(1)_V\times U(1)_A\times Spin(2)_{\text{geo.}}}{\Z_2\times\Z_2},
\end{equation}
where the quotient identifies the $\Z_2$ centers 
of all three groups in the numerator, such that
\begin{align}
\begin{gathered}
    U(1)_{W} = U(1)_V, \\
    U(1)_{S} = \frac{U(1)_A}{\Z_2},\qquad
    SO(2)_{\text{geo.}} = \frac{Spin(2)_{\text{geo.}}}{\Z_2}.
\end{gathered}
\end{align}
The integer pair $(q_{\ell},q_r)$ is thus the charge of
\begin{equation}
    U(1)_L\times U(1)_R \equiv \frac{U(1)_V \times U(1)_A}{\Z_2}
    \label{eq:chiral_symmetry_Dirac},
\end{equation}
and $\Z_2\subset U(1)_{W}$ is therefore identified with fermionic-parity
$\Z_2$ generated by $(-1)^F$.

This symmetry rewrite rearranges the attachments of topological lines on
point operators, without affecting the correlation functions of point
operators on $S^2$. For example, as we see above, a local fermionic
operator necessarily has the $U(1)_{W}$ and $U(1)_{S}$ charges
\begin{equation}
    q_w\in2\Z+1\,,\qquad q_s\in\Z+\frac{1}{2}\,.
\end{equation}
In the bosonic theory~\eqref{eq:compact_scalar}, an operator with this
charge is non-local: its half-integer $U(1)_{S}$ charge shows that it must be
attached to a topological line of $\Z_2\subset U(1)_{W}$, as a
consequence of the mixed 't Hooft anomaly between $U(1)_{W}$ and
$U(1)_{S}$.

The 't Hooft anomaly also gets rewritten by the spin $\theta$-angle and
is now characterized by the 3D fermionic invertible theory
\begin{equation}\label{eq:anomaly_fermion}
    \exp\left\{\frac{i}{4\pi}\int \left(A_L\wedge\d A_L -A_R\wedge\d A_R\right)\right\}
\end{equation}
for background $U(1)_L\times U(1)_R$ gauge field $A_L, A_R$ on a 3D spin manifold.  We can recover the bosonic
anomaly~\eqref{eq:anomaly_boson} if we (illegally) substitute
$A_L=A_W+\frac{1}{2}A_S$ and $A_R=A_W-\frac{1}{2}A_S$. 

The spin $\theta$-angle also affects T-duality. Using Poisson
resummations, one can verify that the torus partition
function~\eqref{eq:torus_Z} is invariant under the composition of
\begin{equation}
    R \to \frac{1}{2R}
\end{equation}
and stacking a phase factor $(-1)^{\Arf(s)}$. This fermionic T-duality
can be proved using the techniques from Appendix~\ref{app:tetrahedron}.
It exchanges $U(1)_V\leftrightarrow U(1)_A$.  Equivalently, it leaves
$U(1)_L$ invariant but conjugates $U(1)_R$.

\subsection{Fermionic description}
\label{sec:fermionic_side}

The fermionic theory~\eqref{eq:compact_scalar_spin} is invariant under stacking $(-1)^{\Arf{s}}$ at the fermionic self-dual radius,
\begin{equation}
    R=\frac{1}{\sqrt{2}}.
\end{equation}
At this special radius, one can verify that the
torus partition function~\eqref{eq:torus_Z} can be reproduced by a 2D free massless Dirac fermion:
\begin{equation}\label{eq:massless_fermion}
    \int\!\!\calD\psi_{\pm}\calD\bar{\psi}_{\pm}
    \exp\left\{-\frac{1}{\pi}\!\int\!\d^2x\Bigl( \bar{\psi}_+\bar{\partial}\psi_+ 
    + \bar{\psi}_-\partial\psi_- \Bigr)\!\right\},
\end{equation}
with left Weyl fermions $\psi_+,\bar{\psi}_+$ and right Weyl fermions
$\psi_-,\bar{\psi}_-$. As this nontrivial fact suggests, the
bosonic path integral~\eqref{eq:compact_scalar_spin} with
$R=\frac{1}{\sqrt{2}}$ and the fermionic path
integral~\eqref{eq:massless_fermion} produce exactly the same fermionic
quantum field theory.

Everything in the bosonic description has a dual in the fermionic
description. The symmetries $U(1)_L$ and $U(1)_R$ rotate the phases of
the left and the right Weyl fermions, respectively.  The fact that these symmetries intersect on the fermionic parity $\Z_2$ symmetry is evident, as is the 't Hooft anomaly of $U(1)_L$ and $U(1)_R$.
The $\Z_2$ symmetry from fermionic T-duality becomes chiral charge conjugation that maps
$\psi_-\leftrightarrow\bar{\psi}_-$ while leaving $\psi_+,\bar{\psi}_+$ invariant. This symmetry ``protects'' the free theory from the Thirring coupling $\bar{\psi}_+\psi_+\bar{\psi}_-\psi_-$.

There is a duality dictionary between all the operators in the two
descriptions. For our purpose, we are interested in the following local
fermionic operators~\cite{Mandelstam:1975hb}:
\begin{equation}\label{eq:cont_duality_dictionary}
\begin{array}{c||c|c}
     & (q_{\ell},q_r) & (q_w,q_s) \\ \hline\hline
    {\color{black!0}\Big|} \psi_+ = \e^{i\theta+i\phi/2} & (1,0) & \makecell{\left(1,\frac{1}{2}\right)} \\ \hline
    {\color{black!0}\Big|} \bar{\psi}_+ = \e^{-i\theta-i\phi/2} & (-1,0) & \left(-1,-\frac{1}{2}\right) \\ \hline
    {\color{black!0}\Big|} \psi_- = \e^{i\theta-i\phi/2} & (0,1) & \left(1,-\frac{1}{2}\right) \\ \hline
    {\color{black!0}\Big|} \bar{\psi}_- = \e^{-i\theta+i\phi/2} & (0,-1) & \left(-1,\frac{1}{2}\right) \\ \hline
\end{array}\ ,
\end{equation}
where $\e^{\pm i\theta}$ denotes the vortex defect operator with winding
number $\pm1$ in the bosonic description. One can verify this dictionary
by comparing correlation functions of these operators on
$S^2\simeq\overline{\C}$. For points on the Riemann sphere
$z_i,w_i\in\overline{\C}$, the fermionic
description~\eqref{eq:massless_fermion} gives
\begin{subequations}\label{eq:general_correlation_fermion}
\begin{gather}
    \left\langle \prod_{i=1}^{N} \psi_+(z_i)\,\bar{\psi}_+(w_i)\right\rangle = \det\left(\frac{1}{z_i-w_j}\right),\\
    \left\langle \prod_{i=1}^{N} \psi_-(z_i)\,\bar{\psi}_-(w_i)\right\rangle = \det\left(\frac{1}{\bar{z}_i-\bar{w}_j}\right),
\end{gather}
\end{subequations}
while the bosonic description~\eqref{eq:compact_scalar_spin} with
$R=\frac{1}{\sqrt{2}}$ gives
\begin{subequations}
\begin{gather}
\begin{split}
    &\:\left\langle \prod_{i=1}^{N} \e^{i\theta(z_i)+i\phi(z_i)/2}\,\e^{-i\theta(w_i)-i\phi(w_i)/2}\right\rangle \\
    =\ &\:\frac{\displaystyle\prod_{1\leq i<j\leq N}(z_i-z_j)(w_j-w_i)}{\displaystyle\prod_{i=1}^N\prod_{j=1}^N (z_i-w_j)}\,,
\end{split}
\\
\begin{split}
    &\:\left\langle \prod_{i=1}^{N} \e^{i\theta(z_i)-i\phi(z_i)/2}\,\e^{-i\theta(w_i)+i\phi(w_i)/2}\right\rangle \\
    =\ &\: \frac{\displaystyle\prod_{1\leq i<j\leq N}(\bar{z}_i-\bar{z}_j)(\bar{w}_j-\bar{w}_i)}{\displaystyle\prod_{i=1}^N\prod_{j=1}^N (\bar{z}_i-\bar{w}_j)}\,.
\end{split}
\end{gather}
\end{subequations}
The results from the two descriptions are actually identical thanks to the
Cauchy determinant formula.

One can generalize this duality by various deformations using the
duality dictionary. For example, relaxing the self-dual condition
$R=\frac{1}{\sqrt{2}}$ in the bosonic
description~\eqref{eq:compact_scalar_spin} corresponds to turning on the
Thirring coupling $\bar{\psi}_+\psi_+\bar{\psi}_-\psi_-$ in the
fermionic description~\eqref{eq:massless_fermion}. Adding a $\cos\phi$
potential in the bosonic description~\eqref{eq:compact_scalar_spin}
corresponds to adding a Dirac mass term
$\bar{\psi}_+\psi_-+\bar{\psi}_-\psi_+$ in the fermionic
description~\eqref{eq:massless_fermion}. Coleman's original
discussion~\cite{Coleman:1974bu} involved both of these deformations,
but we will not consider them in this paper.

\section{Lattice chiral fermion in the 2D Modified Villain Model}
\label{sec:modified_Villain_section}

The 2D modified Villain model furnishes a lattice regularization of the 2D compact scalar~\eqref{eq:compact_scalar} while preserving much of its structure, such as the $U(1)_W\times U(1)_S$ symmetry and its ’t Hooft anomaly.
By the boson-fermion duality reviewed in Section~\ref{sec:Coleman}, it therefore also gives a lattice regularization of the 2D massless Dirac fermion~\eqref{eq:massless_fermion}, up to the spin $\theta$-angle~\eqref{eq:theta_spin}. 

In this section, we explore infinite-lattice correlation functions of fermionic operators in the 2D modified Villain model.
We will not formulate the spin $\theta$-angle on lattice, and consequently our fermion operators come attached to topological lines.
Activating the spin $\theta$-angle removes these topological-line attachments, but does not affect correlation functions on the infinite lattice.

\subsection{2D Modified Villain Model}
\label{sec:the_model}

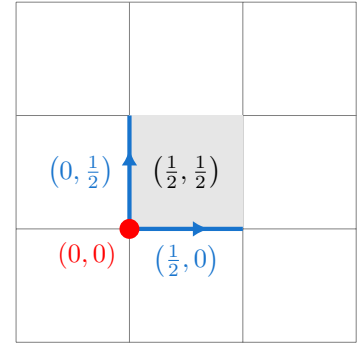
\begin{figure}
\centering
\begin{tikzpicture}[scale=1.5]
\draw [gray, very thin] (0, 0) grid (3, 3);
    \fill[gray!20] (1, 1) rectangle (2, 2); \node[font=\normalsize,
    text=black] at (1.5, 1.5) {$\left(\frac{1}{2},\frac{1}{2}\right)$};
    
    \begin{scope}
      \draw[ DodgerBlue3, ultra thick, postaction={ decorate,
        decoration={ markings, mark=at position 0.7 with
        {\arrow{Latex[length=6pt,width=6pt]}} } } ] (1,1) -- (2,1)
        node[midway, below=2pt, font=\normalsize]
        {$\left(\frac{1}{2},0\right)$};
    
      \draw[ DodgerBlue3, ultra thick, postaction={ decorate,
        decoration={ markings, mark=at position 0.7 with
        {\arrow{Latex[length=6pt,width=6pt]}} } } ] (1,1) -- (1,2)
        node[midway, left=2pt, font=\normalsize]
        {$\left(0,\frac{1}{2}\right)$};
    
    \end{scope}

    \fill[ultra thick, red] (1,1) circle (2.5pt) node[below left,
    xshift=-1pt, yshift=-1pt, font=\normalsize] {$(0,0)$};
    
\end{tikzpicture}
\caption{Lattice sites, links and plaquettes are labeled using the
$\frac{1}{2}$-notation. Links are oriented to point to the right or to
the top.}
\label{fig:1_2_notation}
\end{figure}

We shall work with an infinite 2D square lattice. Let us denote the
collections of sites, links, plaquettes by $\Gamma_0$, $\Gamma_1$, and
$\Gamma_2$, respectively. It is convenient to label the lattice elements
with a ``$\frac{1}{2}$-notation.'' Namely, we identify
\begin{equation}
    \Gamma_0\cup\Gamma_1\cup\Gamma_2\ \simeq\ \frac{1}{2}\Z\oplus\frac{1}{2}\Z
\end{equation}
such that $x\in\Gamma_n$ if $x=(x_1,x_2)$ has exactly $n$ non-integer components.
The orientation of a lattice element aligns with the order of its
non-integer indices. For example, as we show in
Fig.~\ref{fig:1_2_notation}, $(0,\frac{1}{2})$ and $(\frac{1}{2},0)$ are
two links emanating from the site $(0,0)$, while
$(\frac{1}{2},\frac{1}{2})$ is the plaquette attached to the links
$(0,\frac{1}{2})$ and $(\frac{1}{2},0)$. 

We shall use various operators on lattice fields including the lattice
differential $\d$, the lattice codifferential $\delta$, the lattice
laplacian $\Delta\equiv\d\delta+\delta\d$, and the lattice Hodge star
$\star$. They are discretizations of their continuum counterparts and
satisfy similar properties. In
Appendix~\ref{app:LatticeConventions}, we summarize their definitions
and prove their properties using the $\frac{1}{2}$-notation we
introduced above.

The 2D modified Villain model is defined by a lattice path integral
\begin{equation}
    \int\!\calD\varphi\calD n\calD\theta\ \e^{-\calS(\varphi,n,\theta)}
\end{equation}
over three lattice fields
\begin{equation}
    \varphi:\Gamma_0\mapsto\R\,,\quad\!
    n:\Gamma_1\mapsto\Z\,,\quad\!
    \theta:\Gamma_2\mapsto\frac{\R}{2\pi\Z}\,,
\end{equation}
with the path integral measure
\begin{equation}\label{eq:Modified_Villain_measure}
\begin{gathered}
    \int\!\calD\varphi\calD n\calD\theta \equiv\\
    \frac{1}{|\Z|^{|\Gamma_0|}}
    \!\Bigg[\prod_{s\in\Gamma_0} \!\!\int_{\R}\!\!\frac{\d\varphi(s)}{2\pi}\!\Bigg]\!\!
    \Bigg[\prod_{\ell\in\Gamma_1}\!\sum_{n(\ell)\in\Z}\Bigg]\!\!
    \Bigg[\prod_{p\in\Gamma_2}\int_{-\pi}^{\pi}\!\!\frac{\d\theta(p)}{2\pi}\!\Bigg]
\end{gathered}
\end{equation}
and the action (i.e.~the Boltzmann weight)
\begin{equation}\label{eq:Modified_Villain_action}
\begin{gathered}
    \calS (\varphi,n,\theta) \equiv\\
    \frac{R^2}{4\pi}\sum_{\ell\in\Gamma_1} 
    \Bigl[\d \varphi(\ell) + 2\pi n(\ell)\Bigr]^2 
    + i\sum_{p\in\Gamma_2} \theta(p)\d n(p)\,.
\end{gathered}
\end{equation}
We shall exclusively focus on the coupling constant
\begin{equation}
    R = \frac{1}{\sqrt{2}}.
\end{equation}
This lattice path integral has the gauge redundancy
\begin{equation}\label{eq:gauge-redundancy}
    \left\{\begin{aligned}
        \varphi &\to \varphi + 2\pi k\\ 
        n &\to n - \d k
    \end{aligned}\right.\,,
    \qquad k:\Gamma_0\mapsto\Z\,,
\end{equation}
which encodes the compact behavior of $\varphi$. The gauge orbit
contributes a divergence $|\Z|^{|\Gamma_0|}$, which has been canceled
out in Eq.~\eqref{eq:Modified_Villain_measure}.

The lattice field $\varphi$ mod $2\pi$ is the discretization of the
continuum field $\phi$, and $U(1)_{S}$ comes from the $\R$ shift
symmetry of $\varphi$ modulo the gauge redundancy in
Eq.~\eqref{eq:gauge-redundancy}. The lattice field $\theta$ is a
Lagrange multiplier that forces $\d n=0$, and $U(1)_{W}$ is the symmetry
of shifting $\theta$. If we choose to integrate out $\theta$, $U(1)_{W}$
detects the winding number of vortex defects. The modified Villain
lattice model precisely reproduces the mixed 't~Hooft anomaly between
these symmetries on the
lattice~\cite{Sulejmanpasic:2019ytl,Gorantla:2021svj,Berkowitz:2023pnz}.

\subsection{Fermionic operator}
\label{sec:fermionic_operators}

To get fermionic operators on a spacetime lattice, an obstruction to
naively applying the continuum dictionary of
Eq.~\eqref{eq:cont_duality_dictionary} is the fact that $\e^{\pm
\frac{i}{2}\varphi}$ lives on sites while $\e^{\pm i\theta}$ lives on
plaquettes. We thus need a natural operation that coherently associates
a site with a plaquette. This operation is the lattice Hodge star
$\star$ that maps
\begin{equation}
    \Gamma_0\mapsto\Gamma_2\,,\qquad
    \Gamma_1\mapsto\Gamma_1\,,\qquad
    \Gamma_2\mapsto\Gamma_0\,,
\end{equation}
which looks particularly simple in the $\frac{1}{2}$-notation:%
\footnote{ Alternatively, one can choose to set $\star$ to be either
$+(\frac{1}{2},-\frac{1}{2})$, $+(-\frac{1}{2},\frac{1}{2})$, or
$+(-\frac{1}{2},-\frac{1}{2})$. The choice is just a convention. }%
\footnote{ We can also readily see that $\star^2$ is a lattice
translation by $(1,1)$. 
}
\begin{equation}
    \star = +\left(\frac{1}{2},\frac{1}{2}\right).
\end{equation}
Using the lattice Hodge star, we can attempt to define lattice Weyl
operators at a site $x\in\Gamma_0$ through
\begin{subequations}\label{eq:lat_bosonization_naive}
\begin{gather}
    \psi_{\pm}(x) \ \text{``}\!\equiv\!\text{''} \ 
    Z_{\pm} \e^{i\theta(\star x) \pm \frac{i}{2}\varphi(x)}\,,\\
    \bar{\psi}_{\pm}(x) \ \text{``}\!\equiv\!\text{''} \  
    \overline{Z}_{\pm} \e^{-i\theta(\star x) \mp \frac{i}{2}\varphi(x)}\,,
\end{gather}
\end{subequations}
where $Z_{\pm}$ and $\overline{Z}_{\pm}$ are finite normalization
factors, which shall be fixed later in Eq.~\eqref{eq:Z=}. We expect that
in the continuum limit $a\to0$, 
\begin{equation}
    a^{-\frac{1}{2}}\psi_{\pm}
    \quad\text{and}\quad
    a^{-\frac{1}{2}}\bar{\psi}_{\pm}
\end{equation}
approach the continuum Weyl operators of
Section~\ref{sec:fermionic_side}, where $a^{-\frac{1}{2}}$ comes from
the scaling dimension of the continuum Weyl operators. The directional
information contained in $\star$ has the potential to  (and
in fact does) account for the non-scalar nature of lattice Weyl
operators. 

Equation~\eqref{eq:lat_bosonization_naive} on its own cannot be
completely correct because it is not invariant under the lattice gauge
transformation~\eqref{eq:gauge-redundancy}. Restoring gauge invariance
requires attaching $\Z_2\subset U(1)_W$ topological lines to these
operators. On the lattice, a ray (which may look as complicated as a network)
emanating from $x$ can be formally written as a link field
\begin{equation}
    C_x:\Gamma_1\mapsto\Z\,,
\end{equation}
such that its lattice codifferential $\delta C_x:\Gamma_0\mapsto\Z$
satisfies
\begin{equation}
    \forall y\in\Gamma_0,\quad \delta C_x(y) = \delta_{x,y}\,.
\end{equation}
We then define Weyl operators on a lattice ray $C_x$ emanating from $x$,
rather than just $x$ itself: 
\begin{subequations}\label{eq:lat_bosonization}
\begin{align}
    \psi_{\pm}(C_x) \!&\equiv\! Z_{\pm} \e^{i\theta(\star x)} 
    \exp\left\{\!\mp\frac{i}{2}\!\sum_{\ell\in\Gamma_1}\! C_x(\ell)
    \Bigl[\d\varphi(\ell)\!+\!2\pi n(\ell)\!\Bigr] \!\right\} \notag\\ 
    &=\! Z_{\pm} \e^{i\theta(\star x) \pm \frac{i}{2}\varphi(x)} 
    \exp\!\left[\!i\pi\!\! \sum_{\ell\in\Gamma_1}\!C_x(\ell)n(\ell)\!\right]\!, \\
    \bar{\psi}_{\pm}(C_x) \!&\equiv\! \overline{Z}_{\pm} 
    \e^{-i\theta(\star x)} \exp\left\{\!\pm\frac{i}{2}\!\sum_{\ell\in\Gamma_1}\! 
    C_x(\ell)\Bigl[\!\d\varphi(\ell)\!+\!2\pi n(\ell)\!\Bigr] \!\!\right\} \notag\\
    &=\! \overline{Z}_{\pm} \e^{-i\theta(\star x) \mp 
    \frac{i}{2}\varphi(x)} \!\exp\!\left[\!i\pi\!\! 
    \sum_{\ell\in\Gamma_1}\!\!C_x(\ell)n(\ell)\!\right]\!\!,
\end{align}
\end{subequations}
where the two equivalent expressions come from the sum-by-parts
identity~\eqref{eq:SumByParts_1}. These definitions are gauge invariant,
and as we shall see, their dependence on the choice of $C_x$ is
topological.

\subsection{Two-point correlation function}
\label{sec:2pt_fn}

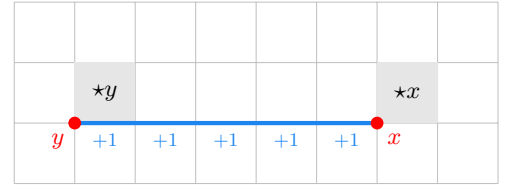
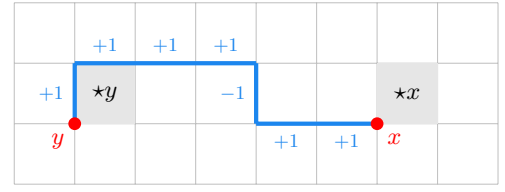
\begin{figure}
\begin{subfigure}[c]{\linewidth}
\centering
\begin{tikzpicture}[scale=0.80]
\draw [gray!50, very thin] (0, 0) grid (8, 3);
    \fill[gray!20] (1, 1) rectangle (2, 2); \node at (1.5, 1.5) {$\star
    y$};
    \fill[gray!20] (6, 1) rectangle (7, 2); \node at (6.5, 1.5) {$\star
    x$};
    \draw[ ultra thick, DodgerBlue2
    ] (1,1) -- (2,1) node[pos=0.5, below=1pt, scale= 0.8]{$+1$} (2,1) --
    (3,1) node[pos=0.5, below=1pt, scale= 0.8]{$+1$} (3,1) -- (4,1)
    node[pos=0.5, below=1pt, scale= 0.8]{$+1$} (4,1) -- (5,1)
    node[pos=0.5, below=1pt, scale= 0.8]{$+1$} (5,1) -- (6,1)
    node[pos=0.5, below=1pt, scale= 0.8]{$+1$};
    \fill[ultra thick, red]
        (1, 1) circle (3pt)  node[below left] {$y$} (6, 1) circle (3pt)
        node[below right] {$x$};
\end{tikzpicture}
\caption{\centering One equivalence class}
\label{fig:correct_branch}
\end{subfigure}

\par\bigskip

\begin{subfigure}[c]{\linewidth}
\centering
\begin{tikzpicture}[scale=0.80]
\draw [gray!50, very thin] (0, 0) grid (8, 3);
    \fill[gray!20] (1, 1) rectangle (2, 2); \node at (1.5, 1.5) {$\star
    y$};
    \fill[gray!20] (6, 1) rectangle (7, 2); \node at (6.5, 1.5) {$\star
    x$};
    \draw[ ultra thick, DodgerBlue2
    ] (1,1) -- (1,2) node[pos=0.5, left=1pt, scale= 0.8]{$+1$} (1,2) --
    (2,2) node[pos=0.5, above=1pt, scale= 0.8]{$+1$} (2,2) -- (3,2)
    node[pos=0.5, above=1pt, scale= 0.8]{$+1$} (3,2) -- (4,2)
    node[pos=0.5, above=1pt, scale= 0.8]{$+1$} (4,2) -- (4,1)
    node[pos=0.5, left=1pt, scale= 0.8]{$-1$} (4,1) -- (5,1)
    node[pos=0.5, below=1pt, scale= 0.8]{$+1$} (5,1) -- (6,1)
    node[pos=0.5, below=1pt, scale= 0.8]{$+1$};
    \fill[ultra thick, red]
        (1, 1) circle (3pt)  node[below left] {$y$} (6, 1) circle (3pt)
        node[below right] {$x$};
\end{tikzpicture}
\caption{\centering The other equivalence class}
\label{fig:wrong_branch}
\end{subfigure}

\caption{Two equivalence classes of lattice paths in the correlation
function
$\left\langle\psi_{\pm}(C_x)\bar{\psi}_{\pm}(C_y)\right\rangle$. A
lattice path $C_{x,y}$ is shown with the indicated values on the blue
links and zero on all other links. $C_{x,y}(\ell)=+1$ indicates that the
path from $y$ to $x$ aligns with the orientation of the link
$\ell\in\Gamma_1$ (given by Fig.~\ref{fig:1_2_notation}), while
$C_{x,y}(\ell)=-1$ indicates the opposite. }
\label{fig:Branches}
\end{figure}


\begin{figure*}[t]
\centering
\begin{tabular}{ccccccc}
\begin{tikzpicture}[scale=0.80]
  \fill[gray!20] (2, 1) rectangle (3, 2); \fill[gray!20] (4, 1)
  rectangle (5, 2);
  \draw[thick, ->] (1, -1.5) -- (2, -1.5) node[right] {$x$};
  \draw[thick, ->] (1, -1.5) -- (1, -0.5) node[above] {$\tau$};
  \draw[ultra thick, DodgerBlue2] (2,1) -- (2,0); \draw[ultra thick,
  DodgerBlue2, dashed] (2,0) -- (2,-1);
  \draw[ultra thick, DodgerBlue2] (4,1) -- (4,0); \draw[ultra thick,
  DodgerBlue2, dashed] (4,0) -- (4,-1);
  \fill[ultra thick, black]
      (2, 1) circle (3pt) node[above right]{$\psi_\pm$} (4, 1) circle
      (3pt) node[above right]{$\psi_\pm$};
\end{tikzpicture}
& \hspace{1em}\raisebox{1.2cm}{\tikz{\draw[thick, ->] (0,0) --
(0.6,0);}}\hspace{-3em} &
\begin{tikzpicture}[scale=0.80]
  \fill[gray!20] (4, 1) rectangle (5, 2); \fill[gray!20] (6, 3)
  rectangle (7, 4);
  \draw[thick, ->,opacity=0] (1, -1.5) -- (2, -1.5) node[right] {$x$};
  \draw[thick, ->,opacity=0] (1, -1.5) -- (1, -0.5) node[above]
  {$\tau$};
  \draw[ultra thick, DodgerBlue2] (6,3) -- (3,3) (3,3) -- (3,0);
  \draw[ultra thick, DodgerBlue2, dashed] (3,0) -- (3,-1);
  \draw[ultra thick, DodgerBlue2] (4,1) -- (4,0); \draw[ultra thick,
  DodgerBlue2, dashed] (4,0) -- (4,-1);
  \fill[ultra thick, black]
      (6, 3) circle (3pt) node[above right]{$\psi_\pm$} (4, 1) circle
      (3pt) node[above right]{$\psi_\pm$};
\end{tikzpicture}
& \hspace{-1em}\raisebox{1.2cm}{\tikz{\draw[thick, ->] (0,0) --
(0.6,0);}}\hspace{-4em} &
\begin{tikzpicture}[scale=0.80]
  \node[red] at (3, 1) {$(-1)\,\times$};
  \fill[gray!20] (4, 1) rectangle (5, 2); \fill[gray!20] (6, 3)
  rectangle (7, 4);
  \draw[thick, ->,opacity=0] (1, -1.5) -- (2, -1.5) node[right] {$x$};
  \draw[thick, ->,opacity=0] (1, -1.5) -- (1, -0.5) node[above]
  {$\tau$};
  \draw[ultra thick, DodgerBlue2] (6,3) -- (5.5,3) (5.5,3) -- (5.5,0);
  \draw[ultra thick, DodgerBlue2, dashed] (5.5,0) -- (5.5,-1);
  \draw[ultra thick, DodgerBlue2] (4,1) -- (4,0); \draw[ultra thick,
  DodgerBlue2, dashed] (4,0) -- (4,-1);
  \fill[ultra thick, black]
      (6, 3) circle (3pt) node[above right]{$\psi_\pm$} (4, 1) circle
      (3pt) node[above right]{$\psi_\pm$};
\end{tikzpicture}
& \raisebox{1.2cm}{\tikz{\draw[thick, ->] (0,0) --
(0.6,0);}}\hspace{-1em} &
\begin{tikzpicture}[scale=0.80]
  \node[red] at (1, 1) {$(-1)\,\times$};
  \fill[gray!20] (2, 1) rectangle (3, 2); \fill[gray!20] (4, 1)
  rectangle (5, 2);
  \draw[thick, ->,opacity=0] (1, -1.5) -- (2, -1.5) node[right] {$x$};
  \draw[thick, ->,opacity=0] (1, -1.5) -- (1, -0.5) node[above]
  {$\tau$};
  \draw[ultra thick, DodgerBlue2] (2,1) -- (2,0); \draw[ultra thick,
  DodgerBlue2, dashed] (2,0) -- (2,-1);
  \draw[ultra thick, DodgerBlue2] (4,1) -- (4,0); \draw[ultra thick,
  DodgerBlue2, dashed] (4,0) -- (4,-1);
  \fill[ultra thick, black]
      (2, 1) circle (3pt) node[above right]{$\psi_\pm$} (4, 1) circle
      (3pt) node[above right]{$\psi_\pm$};
\end{tikzpicture}
\end{tabular}

\caption{The site-plaquette nature of the Weyl operators in combination with the topological line gives rise to the expected fermionic statistics. This illustrates anticommutativity under particle exchange.}
\label{fig:fermi_stats}
\end{figure*}
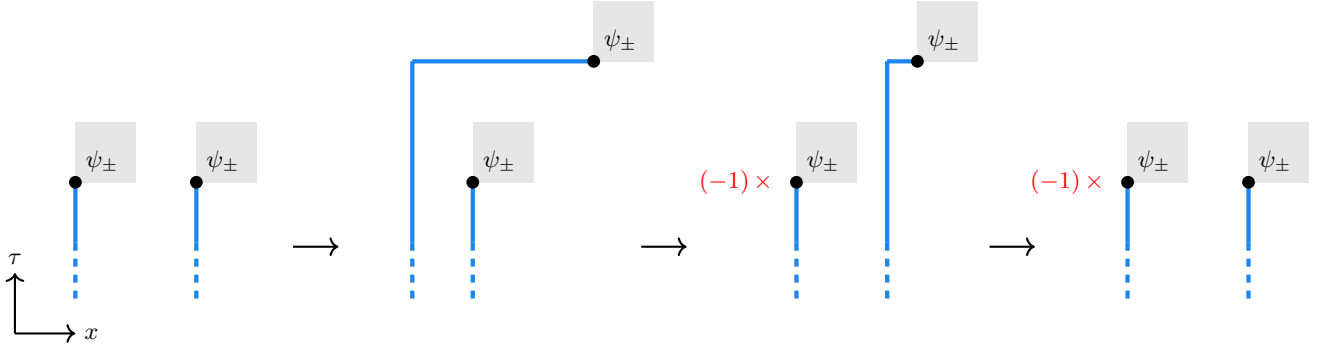


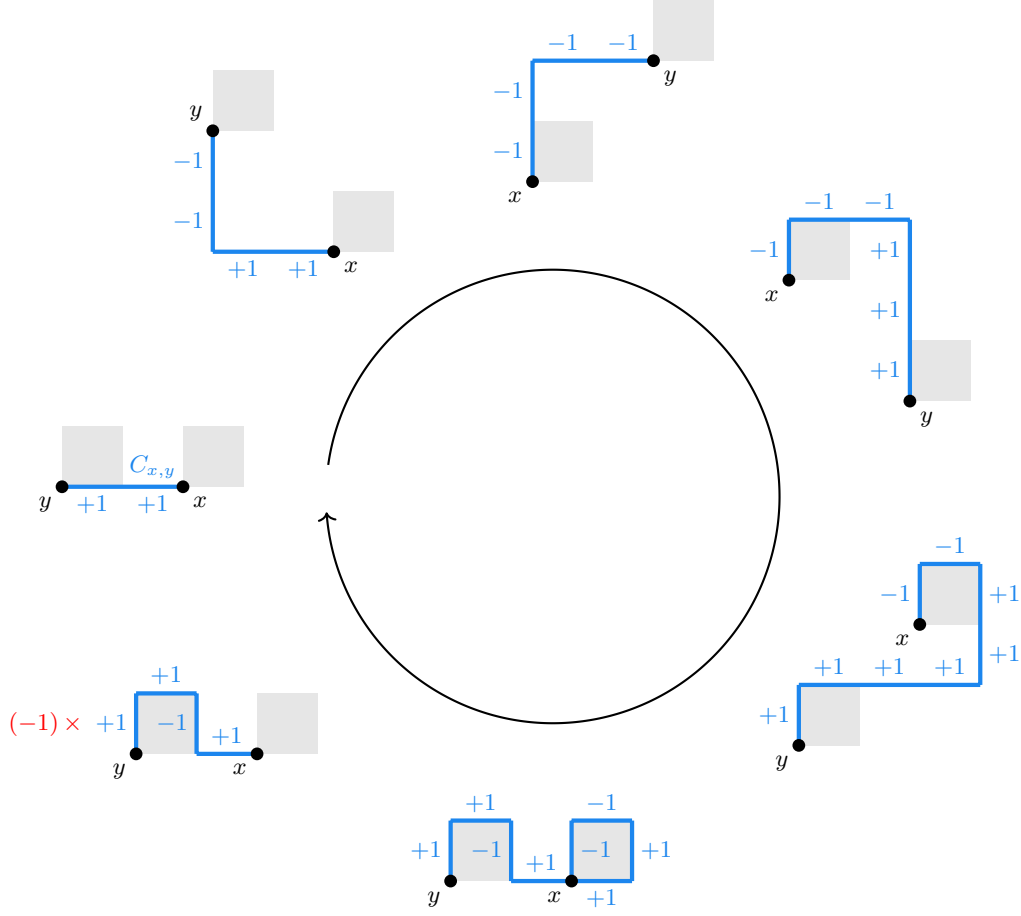
\begin{figure*}[t]
\centering
\begin{tikzpicture}
  \def\R{5}

  \foreach \i/\ang in {1/180, 2/128.57, 3/77.14, 4/25.71, 5/-25.71,
    6/-77.14, 7/-128.57} { \node (fig\i) at (\ang:\R) {}; }

  \node[anchor=center] at (180:\R) {
    \begin{tikzpicture}[scale=0.80]
      \fill[gray!20] (2, 1) rectangle (3, 2); \fill[gray!20] (4, 1)
      rectangle (5, 2); \draw[ultra thick, DodgerBlue2] (2,1) -- (4,1);
      \node[below, DodgerBlue2] at (2.5, 1) {$+1$}; \node[below,
      DodgerBlue2] at (3.5, 1) {$+1$}; \node[above, DodgerBlue2] at
      (3.5, 1) {$C_{x,y}$};
      \fill[ultra thick, black]
          (2, 1) circle (3pt) node[below left]{$y$} (4, 1) circle (3pt)
          node[below right]{$x$};
    \end{tikzpicture}
  };

  \node[anchor=center] at (128.57:\R) {
    \begin{tikzpicture}[scale=0.80]
      \fill[gray!20] (2, 3) rectangle (3, 4); \fill[gray!20] (4, 1)
      rectangle (5, 2);
      \draw[ultra thick, DodgerBlue2]
          (2,1) -- (4,1) (2,1) -- (2,3); \node[left, DodgerBlue2] at (2,
          1.5) {$-1$}; \node[left, DodgerBlue2] at (2, 2.5) {$-1$};
          \node[below, DodgerBlue2] at (2.5, 1) {$+1$}; \node[below,
          DodgerBlue2] at (3.5, 1) {$+1$};
      \fill[ultra thick, black]
          (2, 3) circle (3pt) node[above left]{$y$} (4, 1) circle (3pt)
          node[below right]{$x$};
    \end{tikzpicture}
  };

  \node[anchor=center] at (77.14:\R) {
    \begin{tikzpicture}[scale=0.80]
      \fill[gray!20] (4, 1) rectangle (5, 2); \fill[gray!20] (6, 3)
      rectangle (7, 4);
      \draw[ultra thick, DodgerBlue2]
          (4,1) -- (4,3) (4,3) -- (6,3); \node[left, DodgerBlue2] at (4,
          1.5) {$-1$}; \node[left, DodgerBlue2] at (4, 2.5) {$-1$};
          \node[above, DodgerBlue2] at (4.5, 3) {$-1$}; \node[above,
          DodgerBlue2] at (5.5, 3) {$-1$};
      \fill[ultra thick, black]
          (6, 3) circle (3pt) node[below right]{$y$} (4, 1) circle (3pt)
          node[below left]{$x$};
    \end{tikzpicture}
  };

  \node[anchor=center] at (25.71:\R) {
    \begin{tikzpicture}[scale=0.80]
      \fill[gray!20] (8, 1) rectangle (9, 2); \fill[gray!20] (6, 3)
      rectangle (7, 4);
      \draw[ultra thick, DodgerBlue2]
          (8,1) -- (8,4) (8,4) -- (6,4) (6,4) -- (6,3); \node[left,
          DodgerBlue2] at (8, 1.5) {$+1$}; \node[left, DodgerBlue2] at
          (8, 2.5) {$+1$}; \node[left, DodgerBlue2] at (8, 3.5) {$+1$};
          \node[above, DodgerBlue2] at (7.5, 4) {$-1$}; \node[above,
          DodgerBlue2] at (6.5, 4) {$-1$}; \node[left, DodgerBlue2] at
          (6, 3.5) {$-1$};
      \fill[ultra thick, black]
          (6, 3) circle (3pt) node[below left]{$x$} (8, 1) circle (3pt)
          node[below right]{$y$};
    \end{tikzpicture}
  };

  \node[anchor=center] at (-25.71:\R*1.1) {
    \begin{tikzpicture}[scale=0.80]
      \fill[gray!20] (4, 1) rectangle (5, 2); \fill[gray!20] (6, 3)
      rectangle (7, 4);
      \draw[ultra thick, DodgerBlue2]
          (4,1) -- (4,2) (4,2) -- (7,2) (7,2) -- (7,4) (7,4) -- (6,4)
          (6,4) -- (6,3); \node[left, DodgerBlue2] at (4, 1.5) {$+1$};
          \node[above, DodgerBlue2] at (4.5, 2) {$+1$}; \node[above,
          DodgerBlue2] at (5.5, 2) {$+1$}; \node[above, DodgerBlue2] at
          (6.5, 2) {$+1$}; \node[right, DodgerBlue2] at (7, 2.5) {$+1$};
          \node[right, DodgerBlue2] at (7, 3.5) {$+1$}; \node[above,
          DodgerBlue2] at (6.5, 4) {$-1$}; \node[left, DodgerBlue2] at
          (6, 3.5) {$-1$};
      \fill[ultra thick, black]
          (6, 3) circle (3pt) node[below left]{$x$} (4, 1) circle (3pt)
          node[below left]{$y$};
    \end{tikzpicture}
  };

  \node[anchor=center] at (-86:\R) {
    \begin{tikzpicture}[scale=0.80]
      \fill[gray!20] (2, 1) rectangle (3, 2); \fill[gray!20] (4, 1)
      rectangle (5, 2);
      \draw[ultra thick, DodgerBlue2]
          (2,1) -- (2,2) (2,2) -- (3,2) (3,2) -- (3,1) (3,1) -- (5,1)
          (5,1) -- (5,2) (5,2) -- (4,2) (4,2) -- (4,1); \node[left,
          DodgerBlue2] at (2, 1.5) {$+1$}; \node[above, DodgerBlue2] at
          (2.5, 2) {$+1$}; \node[left, DodgerBlue2] at (3, 1.5) {$-1$};
          \node[above, DodgerBlue2] at (3.5, 1) {$+1$}; \node[below,
          DodgerBlue2] at (4.5, 1) {$+1$}; \node[right, DodgerBlue2] at
          (5, 1.5) {$+1$}; \node[above, DodgerBlue2] at (4.5, 2) {$-1$};
          \node[right, DodgerBlue2] at (4, 1.5) {$-1$};
      \fill[ultra thick, black]
          (2, 1) circle (3pt) node[below left]{$y$} (4, 1) circle (3pt)
          node[below left]{$x$};
    \end{tikzpicture}
  };

  \node[anchor=center] at (-145:\R*1.15) {
    \begin{tikzpicture}[scale=0.80]
      \fill[gray!20] (2, 1) rectangle (3, 2); \fill[gray!20] (4, 1)
      rectangle (5, 2);
      \draw[ultra thick, DodgerBlue2]
          (2,1) -- (2,2) (2,2) -- (3,2) (3,2) -- (3,1) (3,1) -- (4,1);
          \node[left, DodgerBlue2] at (2, 1.5) {$+1$}; \node[above,
          DodgerBlue2] at (2.5, 2) {$+1$}; \node[left, DodgerBlue2] at
          (3, 1.5) {$-1$}; \node[above, DodgerBlue2] at (3.5, 1) {$+1$};
      \fill[ultra thick, black]
          (2, 1) circle (3pt) node[below left]{$y$} (4, 1) circle (3pt)
          node[below left]{$x$}; \node[red] at (0.5, 1.5)
          {$(-1)\,\times$};
    \end{tikzpicture}
  };

  \def\cx{0.5} \def\cy{-0.3}
  
  \draw[thick, ->] plot[variable=\t, domain=172:-176, samples=200,
    smooth] ({\cx + (\R-2)*cos(\t)}, {\cy + (\R-2)*sin(\t)});

\end{tikzpicture}

\caption{When we drag $y$ in a full circle around $x$ for the
correlation function
$\left\langle\psi_{\pm}(C_x)\bar{\psi}_{\pm}(C_y)\right\rangle$, the
final lattice path $C_{x,y}$ lies in the same branch as the initial
lattice path $C_{x,y}$.}
\label{fig:branch_consistency}
\end{figure*}


We now evaluate the two-point correlation functions of the lattice Weyl
operators. All such correlation functions vanish due to symmetry
reasons, except
\begin{equation}\label{eq:two-point_def}
\begin{gathered}
    \left\langle\psi_{\pm}(C_x)\bar{\psi}_{\pm}(C_y)\right\rangle = \\
    \frac{\displaystyle \int\!\!\calD\varphi\calD n\calD\theta\,\e^{-\calS(\varphi,n,\theta)} \psi_{\pm}(C_x)\bar{\psi}_{\pm}(C_y)}{\displaystyle \int\!\!\calD\varphi\calD n\calD\theta\,\e^{-\calS(\varphi,n,\theta)}}.
\end{gathered}
\end{equation}
Due to the topological lines attached to $\psi_{\pm}$, this equation is ambiguous when $x=y$. 
Indeed, one is always free to adjust the contact terms in lattice correlation functions, and we will specify our contact term later.
The composite operator $\psi_{\pm}(C_x)\bar{\psi}_{\pm}(C_y)$  only
depends on $C_y-C_x$. Namely,
\begin{equation}
\begin{gathered}
    \psi_{\pm}(C_x)\bar{\psi}_{\pm}(C_y)
    = Z_{\pm}\overline{Z}_{\pm}\e^{i\theta(\star x)-i\theta(\star y)}\, \times \\ \qquad \qquad
    \exp\left\{\pm\frac{i}{2}\sum_{\ell\in\Gamma_1} 
    C_{x,y}(\ell)\Bigl[\d\varphi(\ell)+2\pi n(\ell)\Bigr] \right\}
\end{gathered}
\end{equation}
where
\begin{equation}\label{eq:2pt_path_C_condition}
    C_{x,y} \equiv C_y - C_x\quad
    \text{s.t.}\quad \delta C_{x,y}(z) = \delta_{y,z} - \delta_{x,z}
\end{equation}
is a lattice path from $y$ to $x$. For fixed $x$ and $y$, all the
lattice paths $C_{x,y}$ fall into two equivalence classes. We show a
representative path $C_{x,y}$ in each class in Fig.~\ref{fig:Branches}.
The correlation function stays the same within each class, and gets
multiplied by $-1$ when jumping from one class to the other. This fact
gives rise to the expected fermionic statistics, as shown in
Fig.~\ref{fig:fermi_stats}.

Consequently, the correlation function, viewed as a function of $x$ and
$y$, is double-valued with two branches differing by $-1$. The two
branches do not mix with each other as we vary $x$ and $y$, 
as shown in Fig.~\ref{fig:branch_consistency}. The role of the spin
$\theta$-angle is to pick one branch consistently to obtain a
single-valued correlation function in the fermionic theory. Let us stick
to the branch represented by Fig.~\ref{fig:correct_branch} (and
Fig.~\ref{fig:branch_consistency}), and work with a single-valued
correlation function
\begin{equation}\label{eq:S(x,y)}
    S_{\pm}(x-y) \equiv \left\langle\psi_{\pm}(C_x)\bar{\psi}_{\pm}(C_y)\right\rangle.
\end{equation}
The other branch represented by Fig.~\ref{fig:wrong_branch} is therefore simply $-S_{\pm}(x-y)$.

We evaluate the correlation function $S_{\pm}(x-y)$ in Appendix~\ref{app:2ptCalc}. The result is expressed in terms of the 2D
lattice Green function $G:\Gamma_0\mapsto\R$ defined by
\begin{equation}
    -\Delta G(x)=\delta_{x,0}\,,\qquad 
    G(0) = 0\,;
\end{equation}
see Appendix~\ref{app:LatticeGreenFunction} for a review. Our
calculation in Appendix~\ref{app:2ptCalc} leads us to, for $x\neq y$,
\begin{align}\label{eq:2pt_with_G}
    S_{\pm}(x-y) = &
    Z_{\pm}\overline{Z}_{\pm} \exp\bigg\{ 2\pi G(x-y) \\ 
    \pm & i\pi \sum_{z\in\Gamma_0} \d C_{x,y}(\star z) 
    \big[G\left(x-z\right) - G\left(y-z\right)\big]\bigg\}.\nonumber
\end{align}
This expression can be simplified when $x-y$ lies along diagonals of the
lattice. For a positive integer $N$, we find
\begin{equation}
\begin{gathered}
    \frac{S_{\pm}(x)}{Z_{\pm}\overline{Z}_{\pm}} = \exp\!\left(\!-\!\sum_{n=1}^{N}\frac{2}{2n\!-\!1}\right)
    \left\{\begin{aligned}
        & 1, && x=(N,-N) \\
        & \!\mp\!i, && x=(N,N) \\
        & \!-\!1, && x=(-N,N) \\
        & \!\pm\!i, && x=(-N,-N)
    \end{aligned}\right.
\end{gathered}.
\end{equation}
Equation~\eqref{eq:2pt_with_G} does not have
illuminating simplified expressions elsewhere.

\begin{figure}
\begin{subfigure}[c]{\linewidth}
    \centering
    \hspace{-1cm}
    \includegraphics[width=0.9\textwidth]{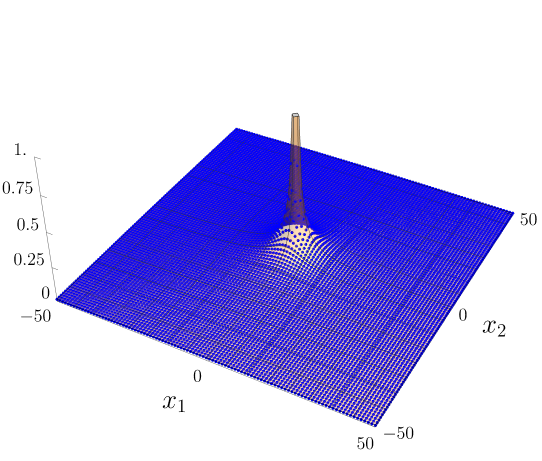}
    \caption{\centering $|S_+(x)|$}
\end{subfigure}

\par\bigskip

\begin{subfigure}[c]{\linewidth}
    \centering
    \hspace{-1cm}
    \includegraphics[width=0.9\textwidth]{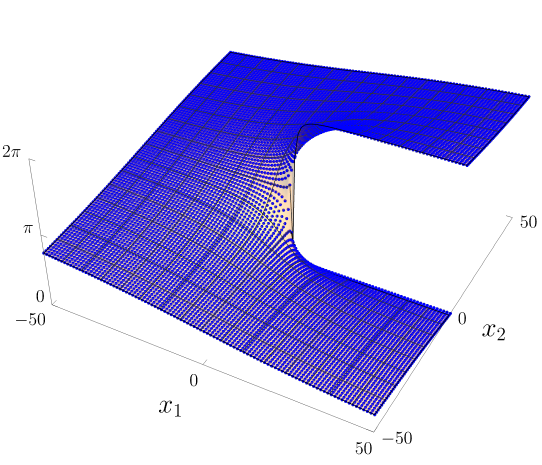}
    \caption{\centering $\arg S_+(x)$}
\end{subfigure}

\caption{Magnitude (top) and phase (bottom) of the correlation function
$S_+(x)$ on a $100\times 100$ lattice. }
\label{fig:PSCor}
\end{figure}

As we prove in Appendix~\ref{app:AsymptoticLatticeWeylCor},
Eq.~\eqref{eq:2pt_with_G} has the large-$|x|$ asymptotic behavior
\begin{equation}\label{eq:correlator_asymptotic}
    S_{\pm}(x) 
    \sim \frac{1}{x_1\pm i x_2} + \calO\left(\frac{1}{|x|^3}\right),
\end{equation}
as long as we fix the renormalization factors such that
\begin{equation}\label{eq:Z=}
    Z_{\pm}\overline{Z}_{\pm} \equiv 2\e^{\gamma_E}\big(1\pm i\big),
\end{equation}
where $\gamma_E$ is the Euler-Mascheroni constant. 
The expansion~\eqref{eq:correlator_asymptotic} implies in the continuum
limit, if we take $a\to 0$ but keep $X\equiv ax$ invariant, we obtain
\begin{equation}\label{eq:continuum_correlation}
    a^{-1}S_{\pm}(x) \sim \frac{1}{X_1\pm iX_2} + \calO(a^2)\,.
\end{equation}
These are exactly the correlation functions in the continuum
theory~\eqref{eq:massless_fermion} of a massless Dirac fermion;
see~Eq.~\eqref{eq:general_correlation_fermion} in the case of $N=2$.
We exhibit a numerical evaluation of $S_+(x)$ in Fig.~\ref{fig:PSCor}.


With such renormalization factors, our correlation functions satisfy the adjoint relation for $x\neq 0$,
\begin{equation}\label{eq:adjoint}
    S_{\mp}(x) = -S_{\pm}^*(-x)\,.
\end{equation}
For the discussion to follow, we want this relation to hold even at $x=0$.
This will work if we choose the contact terms to be 
\begin{equation}
    S_+(0) = c\,,\quad S_-(0) = -c^*\,,\quad c\in\C\,.
\end{equation}
A naive choice is e.g.~$c=Z_+\overline{Z}_+$.  However, other choices may be better.
When $x=(x_1,x_2)\neq(0,0)$, our correlation functions satisfy a $\Z_2\times\Z_2$ symmetry from lattice geometry,
\begin{align}\label{eq:Z2xZ2}
\begin{gathered}
    S_{\pm}(x_1,x_2)
    = -S_{\pm}(-x_1,-x_2) \\
    = \mp i S_{\pm}^*(x_2,x_1) 
    = \pm i S_{\pm}^*(-x_2,-x_1).
\end{gathered}
\end{align}
This symmetry also holds for $x=0$ if we set $c=0$.

\subsection{Non-local lattice Dirac operator}
\label{sec:Nonlocal_Dirac}

In terms of the chiral basis
\begin{equation}
    \gamma^1=\begin{pmatrix}
        0 & 1 \\ 1 & 0
    \end{pmatrix}, \qquad \gamma^2=\begin{pmatrix}
        0 & -i \\ i & 0
    \end{pmatrix},
\end{equation}
the continuum theory~\eqref{eq:massless_fermion} of a massless Dirac
fermion has the hermitian Dirac operator
\begin{equation}\label{eq:continuum_hermitian_dirac}
    i\slashed{\partial} = 2i\left[\begin{array}{cc}
        0 & \displaystyle\frac{\partial}{\partial (x_1\!+\!ix_2)} \\
        \displaystyle\frac{\partial}{\partial (x_1\!-\!ix_2)} & 0
    \end{array}\right].
\end{equation}
The complex differential operators solve the equations
\begin{equation}
    \frac{\partial}{\partial (x_1\mp i x_2)} \frac{1}{x_1\pm i x_2} = \pi\delta^{(2)}(x)\,,
\end{equation}
where $1/(x_1\pm ix_2)$ are the continuum correlation functions as we
see in Eq.~\eqref{eq:continuum_correlation}.

We can reconstruct a hermitian lattice Dirac operator from the modified Villain model, defined as the kernel
\begin{equation}
    \slashed{D}(x,y) \equiv 2i\left[\begin{array}{cc}
        0 & P_+(x-y) \\
        P_-(x-y) & 0
    \end{array}\right] 
\end{equation}
that solve the lattice equations,
\begin{equation}\label{eq:Weyl_eq}
    \sum_{z\in\Gamma_0}P_{\mp}(x-z) \,S_\pm(z-y) = \pi\delta_{x,y}\,.
\end{equation}
This Dirac operator is hermitian because the adjoint relation~\eqref{eq:adjoint} implies
\begin{equation}
    P_{\mp}(x) = -P_{\pm}^*(-x).
\end{equation}
This lattice Dirac operator respects an ``on-site'' chiral symmetry, since it anti-commutes with
\begin{equation}
    \gamma^3\equiv-i\gamma^1\gamma^2=\begin{pmatrix}
        1 & 0 \\ 0 & -1
    \end{pmatrix}.
\end{equation}
We would like to ask the question: how does it evade the Nielsen-Ninomiya theorem?

Using Fourier transformations, we can solve Eq.~\eqref{eq:Weyl_eq} to
obtain
\begin{equation}
    P_-(x) = \int_{T^2}\frac{\d^2p}{(2\pi)^2} \,\e^{i p\cdot x} \frac{\pi}{\widetilde{S}_+(p)},
\end{equation}
where $\widetilde{S}_+(p)$ is the Fourier transform of $S_+(x)$, i.e. 
\begin{equation}\label{eq:S_momentum}
    \widetilde{S}_+(p) = \sum_{x\in\Gamma_0}\e^{-i p\cdot x} S_+(x)\,.
\end{equation}
As we prove in Appendix~\ref{app:analytic_structure}, the large-$|x|$
asymptotic expansion~\eqref{eq:correlator_asymptotic} of $S_+(x)$
implies the small-$|p|$ asymptotic expansion
\begin{equation}\label{eq:pole}
    \widetilde{S}_+(p) \sim \frac{-2\pi i}{p_1\!+\!ip_2} + \calO(1)\,,
\end{equation}
and the vanishing $\calO(|x|^{-2})$ term in
Eq.~\eqref{eq:correlator_asymptotic} implies that the above pole is the
only discontinuity of $\widetilde{S}_+(p)$ on $T^2$. Then according to the
Poincar\'e-Hopf theorem%
\footnote{Note that we can always view a complex scalar field on $T^2$,
such as $\widetilde{S}_+(p)$ and $1/\widetilde{S}_+(p)$, as a 2D real
vector field on $T^2$ through its real and imaginary parts. Thus the
Poincar\'e-Hopf theorem can be applied to this 2D real vector field.}
we reviewed in Section~\ref{sec:Nielsen-Ninomiya},
$\widetilde{S}_+(p)$ has at least one zero, such that the sum of
indices of all zeros equals $+1$ to cancel the pole index $-1$.


\begin{figure}
\begin{subfigure}[c]{\linewidth}
    \centering
    \hspace{-1.5cm}
    \includegraphics[width=0.85\textwidth]{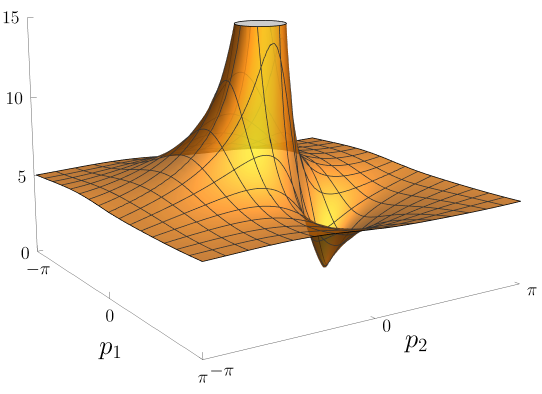}
    \caption{\centering$|\widetilde{S}_+(p)|$}
    \label{fig:MSCor_mag_naive}
\end{subfigure}

\par\bigskip

\begin{subfigure}[c]{\linewidth}
    \centering
    \hspace{-1.5cm}
    \includegraphics[width=0.85\textwidth]{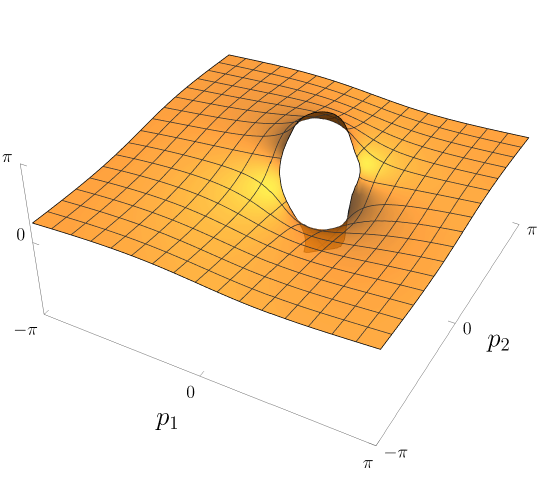}
    \caption{\centering$\arg\widetilde{S}_+(p)$}
    \label{fig:MSCor_arg_naive}
\end{subfigure}

\caption{Magnitude (top) and argument (bottom) of the momentum-space
correlation function $\widetilde{S}_+(p)$, with the naive position-space
contact term $S_+(0) = Z_+\overline{Z}_+$. The pole at $(0,0)$ has index
$-1$ and is the universal feature. For this contact-term choice, the
compensating zero has index $+1$ and lies nearby.}
\label{fig:MSCor_naive}
\end{figure}

\begin{figure}
\begin{subfigure}[c]{\linewidth}
    \centering
    \hspace{-1.5cm}
    \includegraphics[width=0.85\textwidth]{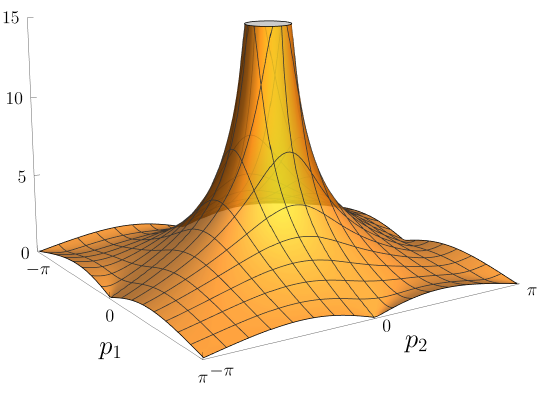}
    \caption{\centering$|\widetilde{S}_+(p)|$}
    \label{fig:MSCor_mag_sym}
\end{subfigure}

\par\bigskip

\begin{subfigure}[c]{\linewidth}
    \centering
    \hspace{-1.5cm}
    \includegraphics[width=0.85\textwidth]{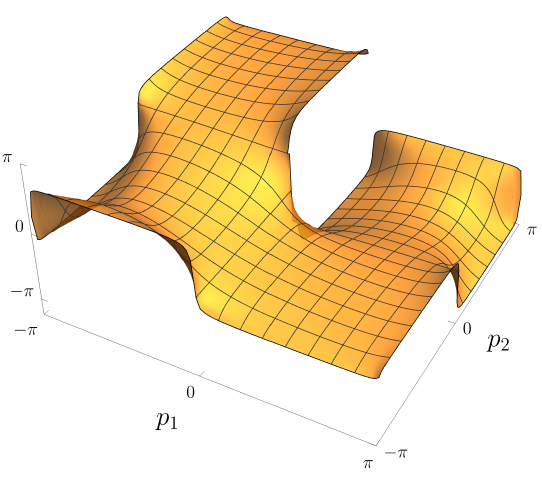}
    \caption{\centering$\arg\widetilde{S}_+(p)$ }
    \label{fig:MSCor_arg_sym}
\end{subfigure}

\caption{A reproduction of Fig.~\ref{fig:MSCor_naive} with the maximally
symmetric choice of the position-space contact term $S_+(0) = 0$. The
pole at $(0,0)$ again has index $-1$. With this contact-term choice, the
compensating zeros instead occur at $(\pi,0),(0,\pi),(\pi,\pi)$ with
indices $+1,+1,-1$, respectively. }
\label{fig:MSCor_sym}
\end{figure}

The precise number and locations of the zeros in $\widetilde{S}_+(p)$
are not protected by the Poincar\'e-Hopf theorem, and depend on the
position-space contact term $S_+(0)$. Changing $S_+(0)$ shifts
$\widetilde{S}_+(p)$ by an overall constant. However, the existence of
at least one zero is not a contact-term artifact: the pole at $p=0$ has
index $-1$, so the total index on $T^2$ can vanish only if the zeros of
$\widetilde{S}_+(p)$ have total index $+1$. 

With the naive contact term
$S_+(0)=Z_+\overline{Z}_+$, as we show in Fig.~\ref{fig:MSCor_naive},
numerically we observe only one zero with index $+1$. Alternatively,
with the maximally symmetric contact term $S_+(0) =0$, the geometric symmetry~\eqref{eq:Z2xZ2} leads to
\begin{equation}\label{eq:sym_S(p)}
\begin{gathered}
    \widetilde{S}_+(p_1,p_2) = - \widetilde{S}_+(-p_1,-p_2)\\
    = i \widetilde{S}_+^*(p_2,p_1) = -i \widetilde{S}_+^*(-p_2,-p_1)\,,
\end{gathered}
\end{equation}
which ensures
\begin{equation}
    \widetilde{S}_+\left(\pi,0\right) = \widetilde{S}_+\left(0,\pi\right) = \widetilde{S}_+\left(\pi,\pi\right) = 0\,.
\end{equation}
And indeed, as we show in Fig.~\ref{fig:MSCor_sym}, they exhaust the
list of zeros we observe numerically, with indices $+1,+1,-1$,
respectively.

Consequently, $1/\widetilde{S}_+(p)$ has exactly one zero, located at
$p=0$, and also robustly has one or more poles elsewhere. The existence of poles in
$1/\widetilde{S}_+(p)$ implies the large-$|x-y|$ asymptotic behavior 
\begin{equation}
    P_{\pm}(x-y) \sim \calO\left(\frac{1}{|x-y|}\right).
\end{equation}
Therefore, the Modified Villain model leads to a non-local lattice Dirac
operator such that
\begin{equation}
    \slashed{D}(x,y) \sim \calO\left(\frac{1}{|x-y|}\right),
\end{equation}
although the model itself is manifestly ultra-local. This is an inevitable
consequence of the Nielsen-Ninomiya theorem.

Before moving on, we note that momentum-space zeros of fermion two-point
functions have played a prominent role in discussions of symmetric mass
generation, where a symmetric gapped phase requires the fermion
propagator to exhibit zeros in place of
poles~\cite{Xu:2021ztz,Lu:2023cev}, and where the significance of such
zeros for constructing lattice chiral gauge theories has been
debated~\cite{Golterman:2023zqf,Golterman:2025boq}. Our results give an
exactly-calculable illustration of the status of propagator zeros: their
total Poincar\'e--Hopf index is fixed topologically, but their number
and locations are not universal, because they can be moved by redefining
the lattice fermion operators by contact terms.

\subsection{Multi-point correlation function}\label{sec:4pt_fn}

Lattice chiral fermions in the 2D Modified Villain model not only propagate
non-locally, but also interact with each other, despite the Gaussian
nature of the model in terms of the original bosonic variables. To see
this, we can consider four-point correlation functions of Weyl
operators. Let us consider the following four sites,
\begin{equation}
\begin{gathered}
    x=(0,L),\quad y=(-L,0),\\
    z=(0,-L),\quad w=(L,0),
\end{gathered}
\end{equation}
for a positive integer $L$. Using the next order of $G(x)$'s asymptotic
expansion of Eq.~\eqref{eq:AsymptoticGreen_NL} and adopting similar
techniques as for the two-point function in
Appendix~\ref{app:AsymptoticLatticeWeylCor}, we can find the following
large-$L$ asymptotic expansion of the magnitudes of these four-point
functions,
\begin{subequations}
\begin{align}
\begin{split}
    &\:\left| \Big\langle \psi_{\pm}(C_x)\psi_{\pm}(C_z)\bar{\psi}_{\pm}(C_y)\bar{\psi}_{\pm}(C_w) \Big\rangle \right| \\
    & \qquad \qquad \qquad \sim \: \frac{1}{L^2} - \frac{5}{24 L^4} + \mathcal{O}\left(\frac{1}{L^6}\right),
\end{split} \\[1em]
\begin{split}
    &\:\Big|\Big\langle \psi_{\pm}(C_x)\bar{\psi}_{\pm}(C_w) \Big\rangle
    \Big\langle \psi_{\pm}(C_z)\bar{\psi}_{\pm}(C_y) \Big\rangle  \\
    &\:- \Big\langle \psi_{\pm}(C_x)\bar{\psi}_{\pm}(C_y) \Big\rangle
    \Big\langle \psi_{\pm}(C_z)\bar{\psi}_{\pm}(C_w) \Big\rangle
    \Big| \\
    & \qquad \qquad \qquad  \sim \:\frac{1}{L^2} - \frac{1}{12 L^4} + \mathcal{O}\left( \frac{1}{L^6} \right).
\end{split}
\end{align}
\end{subequations}
The mismatch of the magnitudes above implies that the connected
four-point function cannot vanish, 
\begin{equation}\label{eq:connected_4pt}
\begin{gathered}
    \Big\langle \psi_{\pm}(C_x) \psi_{\pm}(C_z) \bar{\psi}_{\pm}(C_y) \bar{\psi}_{\pm}(C_w) \Big\rangle_\text{connected} \neq0\,,
\end{gathered}
\end{equation}
which means that the lattice Weyl operators have a nontrivial
finite-spacing two-body interaction. This interaction is nevertheless a
lattice artifact. The factorized contribution scales as $1/L^2$, while
the magnitude mismatch above first appears at order $1/L^4$, consistent
with an interaction that is irrelevant in the continuum limit, where the
theory reduces to the free massless Dirac fermion of
Eq.~\eqref{eq:massless_fermion}.

\section{Outlook}

Bosonization gives a fresh perspective on what the Nielsen-Ninomiya
theorem really says. It only forbids
the existence of a local, translation-invariant, doubler-free Dirac
kernel with the standard chiral structure. 
However, it never forbids a ultra-local lattice model from
having chiral fermions in its long-distance spectrum.
This distinction is
explicitly illustrated by the modified Villain scalar model, where the
microscopic degrees of freedom are bosonic fields, and the action and symmetries are
ultra-local. 
We have shown that the modified-Villain scalar model realizes the physics of a single
massless Dirac fermion without doublers, despite the absence of any microscopic Grassmann
fields.

However, the price of these nice features is that the effective Dirac
operator for the fermionic operators is non-local. This non-locality is
how the Nielsen-Ninomiya theorem is evaded by the bosonized model. Of
course, non-local doubler-free lattice fermions, such as SLAC/Stacey
fermions, have long been known to evade the Nielsen-Ninomiya theorem by
giving up locality~\cite{Drell:1976mj,Rabin:1981nm,Weinstein:1982ht,Stacey:1981ki,Stacey:1983cb,Stacey:1983me,Stacey:1985rqq}. The important
distinction between this older approach and our work here is that our non-local Dirac operator does not act on microscopic Grassmann fields. It is a reconstructed inverse
propagator for certain exotic composite operators in a bosonic lattice
model. While at first glance the non-locality of the Dirac operator may look alarming, the
modified Villain model is completely non-pathological, and is in fact
ultra-local in terms of the microscopic bosonic degrees of freedom.

The locality of the microscopic bosonic variables means that the
non-locality of the reconstructed Dirac operator poses no obstacle to
gauging either the $U(1)_V$ or $U(1)_A$ symmetries: one gauges the symmetries of the local bosonic lattice model
directly, with the non-local Dirac operator playing no role in the
construction. Lattice gauge theories obtained by gauging exact
symmetries of the 2D modified Villain model, including chiral
symmetries, have been explored in e.g.
Refs.~\cite{Berkowitz:2023pnz,DeMarco:2023hoh,Seifnashri:2026ema}.

While bosonization is most well-studied in 2D theories, it is also
believed to be possible in higher dimensions. Unfortunately, above 3D bosonization
remains poorly understood; see e.g.~Refs.~\cite{Burgess:1994tm,Frohlich:1994mj,Chen:2018nog,
Chen:2019wlx,Murugan:2021jwu}. However, in 3D, a massless Dirac fermion
is believed to be dual to an $O(2)$ Wilson-Fisher scalar coupled to a
Chern-Simons
term~\cite{Giombi:2011kc,Aharony:2011jz,Aharony:2012nh,GurAri:2012is,
Aharony:2015mjs,Hsin:2016blu,Karch:2016sxi,Seiberg:2016gmd}. Given the
recent progress in formulating Chern-Simons terms on the lattice in
Refs.~\cite{Jacobson:2023cmr,Jacobson:2024hov}, see also
Refs.~\cite{Chen:2017lkr,Son:2018zja,Chen:2018vmz,Xu:2024hyo,
Peng:2025nfa,Xu:2026ygx,Hatakeyama:2024fzv,Ikeda:2026lyl}, it seems
reasonable to hope that one can use lattice versions of these 3D
dualities to evade the Nielsen-Ninomiya theorem as well. Given our results, it is natural to conjecture that the effective massless Dirac operators
emerging from the bosonic side of these constructions are also non-local.

{\bf Acknowledgements.} 
We are very grateful to Hersh Singh for invaluable communications about the Nielsen-Ninomiya theorem and L\"uscher's reinterpretation of the Ginsparg-Wilson relation, at the SCGP workshop ``Paths to Quantum Field Theory,'' Jun.~22-26, 2026. 
We are also grateful to David B.~Kaplan, Tatsuhiro Misumi, and Srimoyee Sen for helpful remarks.
This work was supported in part by the Simons Foundation through the Collaboration on Confinement and QCD Strings under award number 994302 (A.~C., S.~C., S.~B., M.~N.) and by the National Science Foundation Graduate Research Fellowship under Grant No.~1842400 (M.~N.).

\appendix
\section{Appendix: 2D boson-fermion transformation tetrahedron}
\label{app:tetrahedron}

In this Appendix, we review the topological manipulations that
intertwine 2D bosonic and fermionic
theories~\cite{Tachikawa:2018cer,Thorngren:2018bhj,Karch:2019lnn,Ji:2019ugf}. The
building blocks are the Arf theories. The basic Arf theory is an
invertible topological theory that generates
\begin{equation}
    \Hom\left(\Omega^{Spin}_2,U(1)\right) = \Z_2\,,
\end{equation}
and has the partition function 
\begin{equation}
    (-1)^{\Arf(s)}\,,
\end{equation}
for a spin structure $s$ on a closed 2D oriented manifold. It is the
infinite negative mass limit of a 2D Majorana fermion and describes the
nontrivial topological phase of the Kitaev chain~\cite{Kitaev:2000nmw}.

With an extra $\Z_2$ symmetry, there are two versions of the Arf
invariant, which we shall refer to as the $\widehat{\Arf}$ and the
$\widetilde{\Arf}$ invariants. For a spin structure $s$ and a $\Z_2$
background gauge field $\rho\in H^1(-,\Z_2)$ on a closed oriented
manifold, the $\widehat{\Arf}$ theory has the partition function 
\begin{equation}
    (-1)^{\widehat{\Arf}(s,\rho)} \equiv (-1)^{\Arf(s+\rho)}\,,
\end{equation}
while the $\widetilde{\Arf}$ theory has the partition function 
\begin{equation}
    (-1)^{\widetilde{\Arf}(s,\rho)} \equiv (-1)^{\Arf(s+\rho)+\Arf(s)}\,.
\end{equation}
The three Arf theories, Arf, $\widehat{\Arf}$, and $\widetilde{\Arf}$,
correspond to the three nontrivial elements of
\begin{equation}
    \Hom\left(\Omega^{Spin}_2(B\Z_2),U(1)\right) = \Z_2\times\Z_2\,.
\end{equation}

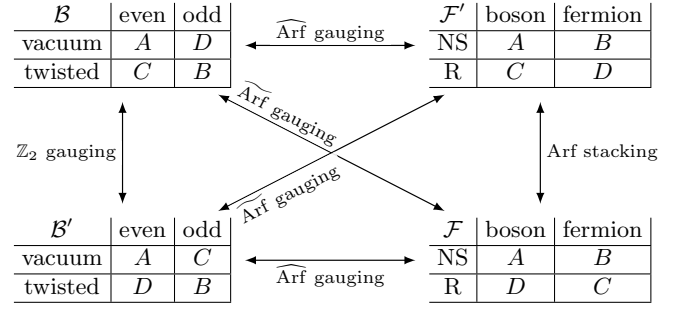
\begin{figure}
\centering
\begin{tikzpicture}[ baseline=(current bounding box.center), transform
    shape, scale=0.98, spectrum/.style={inner sep=0pt},
    manipulation/.style={latex-latex, line width=0.45pt, shorten <=5pt,
    shorten >=5pt}, oplabel/.style={font=\scriptsize, fill=white, fill
    opacity=0.9, text opacity=1, inner xsep=2.2pt, inner ysep=1.4pt} ]
    \node[spectrum] (B) at (-2.82,1.45) {$
        \begin{array}{c|c|c|}
            \calB & \text{even} & \text{odd} \\ \hline
            \text{vacuum} & A & D \\ \hline
            \text{twisted} & C & B \\ \hline
        \end{array}
    $}; \node[spectrum] (Fp) at (2.82,1.45) {$
        \begin{array}{c|c|c|}
            \calF' & \text{boson} & \text{fermion} \\ \hline
            \text{NS} & A & B \\ \hline
            \text{R} & C & D \\ \hline
        \end{array}
    $}; \node[spectrum] (Bp) at (-2.82,-1.45) {$
        \begin{array}{c|c|c|}
            \calB' & \text{even} & \text{odd} \\ \hline
            \text{vacuum} & A & C \\ \hline
            \text{twisted} & D & B \\ \hline
        \end{array}
    $}; \node[spectrum] (F) at (2.82,-1.45) {$
        \begin{array}{c|c|c|}
            \calF & \text{boson} & \text{fermion} \\ \hline
            \text{NS} & A & B \\ \hline
            \text{R} & D & C \\ \hline
        \end{array}
    $};

    \draw[manipulation] (B) -- node[oplabel, above] {$\widehat{\Arf}$
    gauging} (Fp); \draw[manipulation] (Bp) -- node[oplabel, below]
    {$\widehat{\Arf}$ gauging} (F); \draw[manipulation] (B) --
    node[oplabel, left] { $\Z_2$ gauging} (Bp); \draw[manipulation] (Fp)
    -- node[oplabel, right] {Arf stacking} (F); \draw[manipulation] (B)
    -- node[oplabel, sloped, above, pos=0.30] {$\widetilde{\Arf}$
    gauging} (F); \draw[manipulation] (Bp) -- node[oplabel, sloped,
    below, pos=0.30] {$\widetilde{\Arf}$ gauging} (Fp);
\end{tikzpicture}
\caption{The tetrahedron of topological manipulations connecting bosonic
theories $\calB$, $\calB'$ and fermionic theories $\calF$,
$\calF'$~\cite{Tachikawa:2018cer,Thorngren:2018bhj,Karch:2019lnn,Ji:2019ugf}.}
\label{fig:shuffle}
\end{figure}

As illustrated in Fig.~\ref{fig:shuffle}, using Arf theories, we can
construct a tetrahedron of topological manipulations that connect four
related 2D theories. There $\calB$ and $\calB'$ are 2D bosonic theories
that have a non-anomalous $\Z_2$ symmetry. Their operators may be even
or odd under $\Z_2$, and may be attached to a $\Z_2$ topological defect
line (so that they live in the $\Z_2$-twisted sector) or not (so that
they live in the vacuum sector). $\calF$ and $\calF'$ are 2D fermionic
theories that have non-anomalous fermionic parity $(-1)^F$. Their
operators may be bosonic or fermionic, and may be attached to a $(-1)^F$
topological defect line (so that they live in the Ramond sector) or not
(so that they live in the Neveu-Schwarz sector). In
Fig.~\ref{fig:shuffle}, sectors with the same label are isomorphic.

The topological manipulations relating the four theories are as
follows~\cite{Tachikawa:2018cer,Thorngren:2018bhj,Karch:2019lnn,Ji:2019ugf}. Let
$Z$ denote the partition function on a closed oriented manifold of genus
$g$. Then the transformation $\calB\rightleftharpoons\calB'$ is realized
by gauging $\Z_2$, e.g.
\begin{equation}
    Z_{\mathcal{B}'}(\rho') = \frac{1}{2^{g}}\sum_{\rho} 
    (-1)^{\int \rho' \cup \rho } Z_{\mathcal{B}}(\rho).
\end{equation}
The transformation $\calF\rightleftharpoons\calF'$ is realized by
stacking $\calF$ with the basic Arf theory, e.g.
\begin{equation}
    Z_{\mathcal{F}'}(s) = (-1)^{\Arf(s)} Z_{\mathcal{F}}(s).
\end{equation}
The transformations $\calB\rightleftharpoons\calF$ and
$\calB'\rightleftharpoons\calF'$ are realized by gauging $\Z_2$ or
$(-1)^F$ with the $\widetilde{\Arf}$ theory, e.g.
\begin{subequations}\label{eq:F_vs_B_partition}
\begin{gather}
    Z_{\mathcal{F}}(s) = \frac{1}{2^{g}}\sum_{\rho} (-1)^{\widetilde{\Arf}(s,\rho)}
    Z_{\mathcal{B}}(\rho),\\
    Z_{\mathcal{B}}(\rho) = \frac{1}{2^{g}}\sum_{s} (-1)^{\widetilde{\Arf}(s,\rho)}
    Z_{\mathcal{F}}(s).\label{eq:F->B}
\end{gather}
\end{subequations}
The transformations $\calB\rightleftharpoons\calF'$ and
$\calB'\rightleftharpoons\calF$ are realized by gauging $\Z_2$ or
$(-1)^F$ with the $\widehat{\Arf}$ theory, e.g.
\begin{subequations}
\begin{gather}
    Z_{\mathcal{F}'}(s) = \frac{1}{2^{g}}\sum_{\rho} (-1)^{\widehat{\Arf}(s,\rho)}
    Z_{\mathcal{B}}(\rho),\\
    Z_{\mathcal{B}}(\rho) = \frac{1}{2^{g}}\sum_{s} (-1)^{\widehat{\Arf}(s,\rho)}
    Z_{\mathcal{F}'}(s).
\end{gather}
\end{subequations}
Using two fundamental properties of the Arf invariant,
\begin{subequations}\label{eq:Arf_properties}
\begin{gather}
    (-1)^{\widetilde{\Arf}(s,\rho_1+\rho_2) + \widetilde{\Arf}(s,\rho_1) + \widetilde{\Arf}(s,\rho_2) + \int\rho_1\cup\rho_2} = 1,\\
    \frac{1}{2^g}\sum_s (-1)^{\Arf(s)} = 1\,,
\end{gather}
\end{subequations}
one can prove the commutativity of the transformation tetrahedron in
Fig.~\ref{fig:shuffle}.

Now let us apply the topological manipulation $\calB\to\calF'$ to the
bosonic theory~\eqref{eq:compact_scalar} of
Section~\ref{sec:bosonic_side} with respect to $\Z_2\subset U(1)_{W}$.
This amounts to an insertion of
\begin{equation}
\begin{split}
    \frac{1}{2^g}\sum_{\rho}(-1)^{\widehat{\Arf}(s,\rho) + \int\left[\frac{\d\phi}{2\pi}\right]_2\cup \rho}
\end{split}
\end{equation}
in the path integral~\eqref{eq:compact_scalar}. Using the fundamental
properties~\eqref{eq:Arf_properties} of the Arf invariant, we have
\begin{equation}
\begin{split}
    &\: \frac{1}{2^g}\sum_{\rho}(-1)^{\widehat{\Arf}(s,\rho) + \int\left[\frac{\d\phi}{2\pi}\right]_2\cup \rho} \\
    =&\: \frac{1}{2^g}\sum_{\rho}(-1)^{\widetilde{\Arf}\left(s,\left[\frac{\d\phi}{2\pi}\right]_2\right) + \Arf\left(s+\left[\frac{\d\phi}{2\pi}\right]_2+\rho\right)} \\
    =&\: (-1)^{\widetilde{\Arf}\left(s,\left[\frac{\d\phi}{2\pi}\right]_2\right)}\,,
\end{split}
\end{equation}
which is precisely the spin $\theta$-angle~\eqref{eq:theta_spin} of
Section~\ref{sec:bosonic_side}. Using the transformation tetrahedron,
one can easily prove the fermionic T-duality we discuss around the end of
Section~\ref{sec:bosonic_side}.

\section{Appendix: 2D lattice conventions}

In this appendix, we review fundamental notions about the 2D square
lattice.

\subsection{Lattice differential operators}
\label{app:LatticeConventions}

In this subsection, we review the discrete differential operators on a
2D lattice. We use $\Gamma_0$, $\Gamma_1$, and $\Gamma_2$ to denote the
collection of sites, links, and plaquettes, respectively. As we
explained in Section~\ref{sec:the_model}, it is convenient to label the
lattice elements with the $\frac{1}{2}$-notation,
\begin{equation}
    \Gamma_0\cup\Gamma_1\cup\Gamma_2\ \simeq\ \frac{1}{2}\Z\oplus\frac{1}{2}\Z
\end{equation}
such that $x\in\Gamma_n$ if $x$ has exactly $n$ non-integer components.
The orientation of a lattice element aligns with the order of its
non-integer indices. See Fig.~\ref{fig:1_2_notation} for a concrete
example.

\begin{figure}[t]
\centering
\begin{tikzpicture}[scale=1.6]
    \draw [gray, very thin] (0, 0) grid (2, 2);

    \fill[gray!20] (0, 0) rectangle (2, 2); \node[font=\normalsize,
    text=black] at (1, 1) {$\d\beta\left(x_1,x_2\right)$};

    \draw[DodgerBlue3, ultra thick, postaction={decorate,
        decoration={markings, mark=at position 0.55 with
        {\arrow{Latex[length=6pt,width=6pt]}}}}] (0,0) -- (2,0)
        node[midway, below=2pt, font=\normalsize]
        {$\beta\left(x_1,x_2\!-\!\tfrac{1}{2}\right)$};

    \draw[DodgerBlue3, ultra thick, postaction={decorate,
        decoration={markings, mark=at position 0.55 with
        {\arrow{Latex[length=6pt,width=6pt]}}}}] (2,0) -- (2,2)
        node[midway, right=2pt, font=\normalsize]
        {$\beta\left(x_1\!+\!\tfrac{1}{2},x_2\right)$};

    \draw[DodgerBlue3, ultra thick, postaction={decorate,
        decoration={markings, mark=at position 0.55 with
        {\arrow{Latex[length=6pt,width=6pt]}}}}] (0,2) -- (2,2)
        node[midway, above=2pt, font=\normalsize]
        {$\beta\left(x_1,x_2\!+\!\tfrac{1}{2}\right)$};

    \draw[DodgerBlue3, ultra thick, postaction={decorate,
        decoration={markings, mark=at position 0.55 with
        {\arrow{Latex[length=6pt,width=6pt]}}}}] (0,0) -- (0,2)
        node[midway, left=2pt, font=\normalsize]
        {$\beta\left(x_1\!-\!\tfrac{1}{2},x_2\right)$};

    \fill[red] (0,0) circle (2.5pt); \fill[red] (2,0) circle (2.5pt);
    \fill[red] (2,2) circle (2.5pt); \fill[red] (0,2) circle (2.5pt);

\end{tikzpicture}
\caption{The lattice differential~\eqref{eq:d_beta_1} of a link field.}
\label{fig:LatticeDifferentialPlaquette}
\end{figure}
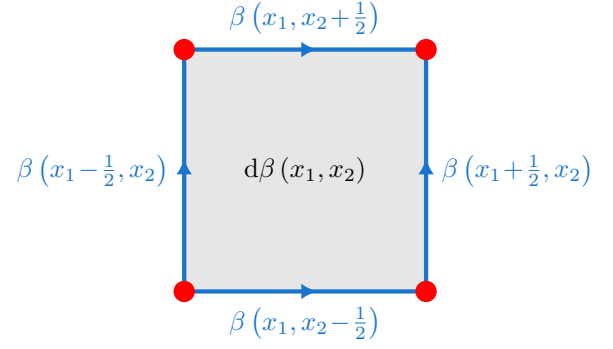

The lattice differential $\d$ maps a $\Gamma_n$ field to a
$\Gamma_{n+1}$ field. For a site field $\alpha$, its differential
$\d\alpha$ is a link field such that $\forall(x_1,x_2)\in\Gamma_1$,
\begin{equation}
\begin{gathered}
    \d\alpha\left(x_1,x_2\right) \equiv \\
    \begin{cases}
        \alpha\left(x_1\!+\!\tfrac{1}{2},x_2\right) - \alpha\left(x_1\!-\!\tfrac{1}{2},x_2\right)\!, & x_1\notin\Z \\
        \alpha\left(x_1,x_2\!+\!\tfrac{1}{2}\right) - \alpha\left(x_1,x_2\!-\!\tfrac{1}{2}\right)\!, & x_2\notin\Z
    \end{cases}
\end{gathered}
\end{equation}
For a link field $\beta$, its differential $\d\beta$ is a plaquette
field such that $\forall(x_1,x_2)\in\Gamma_2$,
\begin{equation}\label{eq:d_beta_1}
\begin{gathered}
    \d\beta\left(x_1,x_2\right) \equiv \\
    \beta\left(x_1\!+\!\tfrac{1}{2},x_2\right) - \beta\left(x_1\!-\!\tfrac{1}{2},x_2\right) \\
    - \,\beta\left(x_1,x_2\!+\!\tfrac{1}{2}\right) + \beta\left(x_1,x_2\!-\!\tfrac{1}{2}\right),
\end{gathered}
\end{equation}
as illustrated by Fig.~\ref{fig:LatticeDifferentialPlaquette}. For a
plaquette field $\gamma$, we have $\d\gamma=0$ due to dimension
overflow.

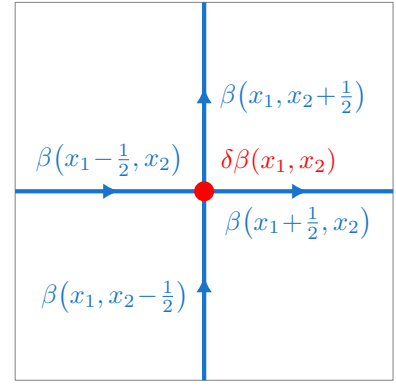
\begin{figure}[t]
\centering
\begin{tikzpicture}[scale=2.5]
    \draw [gray, very thin] (-1, -1) grid (1, 1);

    \coordinate (Center) at (0, 0); \coordinate (S) at (0, -1);
    \coordinate (E) at (1, 0); \coordinate (N) at (0, 1); \coordinate
    (W) at (-1, 0);

    \draw[DodgerBlue3, ultra thick, postaction={decorate,
        decoration={markings, mark=at position 0.55 with
        {\arrow{Latex[length=6pt,width=6pt]}}}}] (S) -- (Center)
        node[midway, left=2pt, yshift=-4pt, font=\normalsize]
        {$\beta\!\left(x_1, x_2\!-\!\tfrac{1}{2}\right)$};
        \draw[DodgerBlue3, ultra thick, postaction={decorate,
        decoration={markings, mark=at position 0.55 with
        {\arrow{Latex[length=6pt,width=6pt]}}}}] (Center) -- (E)
        node[midway, below=2pt, font=\normalsize]
        {$\beta\!\left(x_1\!+\!\tfrac{1}{2}, x_2\right)$};
        \draw[DodgerBlue3, ultra thick, postaction={decorate,
        decoration={markings, mark=at position 0.55 with
        {\arrow{Latex[length=6pt,width=6pt]}}}}] (Center) -- (N)
        node[midway, right=2pt, font=\normalsize] {$\beta\!\left(x_1,
        x_2\!+\!\tfrac{1}{2}\right)$}; \draw[DodgerBlue3, ultra thick,
        postaction={decorate, decoration={markings, mark=at position
        0.55 with {\arrow{Latex[length=6pt,width=6pt]}}}}] (W) --
        (Center) node[midway, above=2pt, font=\normalsize]
        {$\beta\!\left(x_1\!-\!\tfrac{1}{2}, x_2\right)$};

    \fill[red] (Center) circle (1.5pt) node[above right, xshift=3pt,
    yshift=3pt, font=\normalsize] {$\delta\beta\!\left(x_1,
    x_2\right)$};
   
\end{tikzpicture}
\caption{The lattice codifferential~\eqref{eq:d_beta_2} of a link
field.}
\label{fig:LatticeCodifferentialSite}
\end{figure}

The lattice codifferential $\delta$ maps a $\Gamma_n$ field to a
$\Gamma_{n-1}$ field. For a plaquette field $\gamma$, its codifferential
$\delta\gamma$ is a link field such that $\forall(x_1,x_2)\in\Gamma_1$,
\begin{equation}
\begin{gathered}
    \delta\gamma\left(x_1,x_2\right) \equiv \\
    \begin{cases}
        \gamma\left(x_1\!+\!\tfrac{1}{2},x_2\right) - \gamma\left(x_1\!-\!\tfrac{1}{2},x_2\right)\!, & x_1\in\Z \\
        -\gamma\left(x_1,x_2\!+\!\tfrac{1}{2}\right) + \gamma\left(x_1,x_2\!-\!\tfrac{1}{2}\right)\!, & x_2\in\Z
    \end{cases}
\end{gathered}
\end{equation}
For a link field $\beta$, its codifferential $\delta\beta$ is a site
field such that $\forall(x_1,x_2)\in\Gamma_0$,
\begin{equation}\label{eq:d_beta_2}
\begin{gathered}
    \delta\beta\left(x_1,x_2\right) \equiv \\
    \beta\left(x_1\!+\!\tfrac{1}{2},x_2\right) - \beta\left(x_1\!-\!\tfrac{1}{2},x_2\right) \\
    + \,\beta\left(x_1,x_2\!+\!\tfrac{1}{2}\right)- \beta\left(x_1,x_2\!-\!\tfrac{1}{2}\right),
\end{gathered}
\end{equation}
as shown in Fig.~\ref{fig:LatticeCodifferentialSite}. For a site field
$\alpha$, we have $\delta\alpha=0$ due to dimension underflow.


\tikzmath{ \w = 3; \u = 4.75; }

Using these definitions, we can show that $\d^2=0$ and $\delta^2=0$, as
well as the sum-by-parts identities,
\begin{subequations} \label{eq:SumByParts}
\begin{gather}
    \sum_{x\in\Gamma_0}\alpha(x)\,\delta\beta(x) + \sum_{y\in\Gamma_1}\d\alpha(y)\,\beta(y) = 0 \,,\label{eq:SumByParts_1}\\
    \sum_{y\in\Gamma_1}\beta(y)\,\delta\gamma(y) + \sum_{z\in\Gamma_2}\d\beta(z)\,\gamma(z) = 0 \,.
\end{gather}
\end{subequations}
The lattice Laplacian $\Delta$ is defined by
\begin{equation}
    \Delta \equiv \d\delta + \delta\d
\end{equation}
and maps a $\Gamma_n$ field to a $\Gamma_n$ field. From this definition,
we can show that no matter if $\sigma$ is a site, link, or plaquette
field, we always have
\begin{equation} \label{eq:DeltaSite}
\begin{gathered}
    \Delta\sigma(x_1,x_2) = \\
    \sigma(x_1\!+\!1,x_2) + \sigma(x_1\!-\!1,x_2) + \sigma(x_1,x_2
    \!+\!1) \\
    + \,\sigma(x_1,x_2\!-\!1) - 4\sigma(x_1,x_2)
\end{gathered}
\end{equation}
for $(x_1,x_2)\in\Gamma_0$, $\in\Gamma_1$ or $\in\Gamma_2$,
respectively.

\subsection{Lattice Green function}
\label{app:LatticeGreenFunction}

In this subsection, we review the lattice Green function $G:\Z^2 \mapsto
\R$ that solves the 2D lattice Poisson equation, 
\begin{equation}\label{eq:PoissonEquation}
    -\Delta G(x) = \delta_{x, 0}\,.
\end{equation}
If $G(x)$ is a solution, so is $G(x)+c$. We fix this constant by
requiring
\begin{equation}\label{eq:G_0_0_Convention}
    G(0) = 0\,.
\end{equation}
The Poisson equation~\eqref{eq:PoissonEquation} and the
convention~\eqref{eq:G_0_0_Convention} determine a unique $G(x)$.

$G(x)$ does not have a particularly illuminating compact expression,
except for special $x$. From the $D_8$ symmetry of the lattice, we can
immediately find
\begin{equation}
    G(1,0) = G(-1,0) = G(0,1) = G(0,-1) = -\frac{1}{4}.
\end{equation}
Since Eq.~\eqref{eq:DeltaSite} implies
\begin{equation}\label{eq:Delta_momentum}
    -\Delta\,\e^{i p\cdot x} = (4-2\cos{p_1}-2\cos{p_2})\,\e^{i p\cdot x} \,,
\end{equation}
we can solve $G(x)$ by the Fourier transformation,
\begin{equation}\label{eq:GreenFunctionFourier}
    G(x) =
    \int_{-\pi}^{\pi} \int_{-\pi}^{\pi} 
    \frac{\d^2p}{(2\pi)^2} \,
    \frac{\e^{i p\cdot x} - 1}{ 4 - 2\cos{p_1} - 2 \cos{p_2} }\,.
\end{equation}
The subtraction $-1$ regularizes the IR divergence at $p=0$, which is
mathematically packaged into the definition of
$1/(4-2\cos{p_1}-2\cos{p_2})$ as a distribution (i.e.~generalized
function) on $T^2$. At special positions with $|x_1|=|x_2|=N$, the
integral~\eqref{eq:GreenFunctionFourier} can be evaluated to
\begin{equation}\label{eq:G_diagonal}
\begin{split}
    G(x) &\:= -\frac{1}{\pi}\sum_{n=1}^N \frac{1}{2n-1} \\
    &\:= -\frac{1}{2\pi}\left[\psi\!\left(\!N\!+\!\frac{1}{2}\right)+\gamma_E + 2\ln2\right],
\end{split}
\end{equation}
where $\psi(z)$ is the digamma function and $\gamma_E$ is the
Euler-Mascheroni constant. Using the Poisson
equation~\eqref{eq:PoissonEquation} and the lattice $D_8$ symmetry, we
can construct $G(x)$ for all remaining $x$ algebraically from the
diagonal values of Eq.~\eqref{eq:G_diagonal}. This algebraic method is
sufficient for numerical purposes, and also leads us to an interesting
property: $G(x)\in\Q+\frac{1}{\pi}\Q$.

We are also interested in the large-$|x|$ asymptotic expansion of
$G(x)$. The lattice Laplacian can be expanded in terms of continuum
differential operators,
\begin{equation}\label{eq:Laplacian_expansion}
    \Delta = \sum_{n=1}^{\infty} \frac{2}{(2n)!} \left( \frac{\partial^{2n}}{\partial x_1^{2n}} + \frac{\partial^{2n}}{\partial x_2^{2n}} \right).
\end{equation}
This is just a Taylor expansion in momentum space;
c.f.~Eq.~\eqref{eq:Delta_momentum}. We can then solve differential
equations order by order to find the large-$|x|$ asymptotic expansion.
The $n$th term Eq.~\eqref{eq:Laplacian_expansion} gives a contribution
of order
\begin{equation}
    \mathcal{O}\left(\frac{1}{|x|^{2n-2}}\right) \,.
\end{equation}
The leading $n=1$ term is just the continuum Laplacian and we thus have
\begin{equation}\label{eq:AsymptoticGreen}
    G(x) \sim -\frac{\ln |x|}{2\pi} 
    - \frac{\gamma_E+\frac{3}{2}\ln2}{2\pi} + \mathcal{O}\left(\frac{1}{|x|^2}\right),
\end{equation}
where the constant term comes from the asymptotic expansion of
Eq.~\eqref{eq:G_diagonal}. While Eq.~\eqref{eq:AsymptoticGreen} is
isotropic, the next order begins to have angular dependence:
\begin{equation}\label{eq:AsymptoticGreen_NL}
    G(x) \sim
    -\frac{\ln r}{2\pi} 
    - \frac{\gamma_E+\frac{3}{2}\ln2}{2\pi} 
    +\frac{\cos4\theta}{24\pi \, r^2}
    + \mathcal{O}\left(\frac{1}{r^4}\right)
\end{equation}
for $x=(r\cos\theta,r\sin\theta)$.

\section{Appendix: Calculation details in the 2D modified Villain model}

In this appendix, we present calculation details of our key results in
the 2D modified Villain model.

\subsection{Two-point correlation function}
\label{app:2ptCalc}

In this subsection, we evaluate the two-point correlation function
$S_{\pm}(x)$. Based on our discussion in Section~\ref{sec:2pt_fn} before
Eq.~\eqref{eq:S(x,y)}, we have
\begin{equation}\label{eq:S_calc}
    S_{\pm}(x-y) = \frac{\displaystyle \int\!\!\calD\varphi\calD n\calD\theta\,\e^{-\calS(\varphi,n,\theta)} \psi_{\pm}(C_x)\bar{\psi}_{\pm}(C_y)}{\displaystyle \int\!\!\calD\varphi\calD n\calD\theta\,\e^{-\calS(\varphi,n,\theta)}}
\end{equation}
with the equivalence class of paths $C_{x,y}\equiv C_y-C_x$ represented
by Fig.~\ref{fig:correct_branch} (and
Fig.~\ref{fig:branch_consistency}).

First let us integrate out $\theta$ in Eq.~\eqref{eq:S_calc}. We are
then left with a constraint
\begin{equation}
    \d n = 0
\end{equation}
for the denominator and a constraint
\begin{equation}\label{eq:dn=}
    \forall p\in\Gamma_2\,,\quad
    \d n(p) = \delta_{p,\star x} - \delta_{p, \star y}
\end{equation}
for the numerator. Then we can gauge fix to eliminate $n$. Different IR
regularizations should lead to the same result in the thermodynamic
limit, so we can simply imagine our spacetime as the large-size limit of
$S^2$. Because $\pi_1(S^2)=0$, the link field $n$ cannot have nontrivial
global holonomies. Hence for both the denominator and the numerator of
Eq.~\eqref{eq:S_calc}, the constraints above leave us with only one
gauge equivalence class. Eliminating the path integral over $n$ by gauge
fixing, we obtain
\begin{widetext}
\begin{equation}\label{eq:S_intermediate}
    \frac{S_{\pm}(x-y)}{Z_{\pm}\overline{Z}_{\pm}} = 
    \frac{\displaystyle \prod_{s\in\Gamma_0} \int_{\R}\frac{\d\varphi(s)}{2\pi} \exp\left\{-\frac{1}{8\pi}\sum_{\ell\in\Gamma_1} 
    \Bigl[\d \varphi(\ell) + 2\pi n_{x,y}(\ell)\Bigr]^2
    \pm\frac{i}{2}\sum_{\ell\in\Gamma_1} 
    C_{x,y}(\ell)\Bigl[\d\varphi(\ell)+2\pi n_{x,y}(\ell)\Bigr] \right\}}
    {\displaystyle \prod_{s\in\Gamma_0} \int_{\R}\frac{\d\varphi(s)}{2\pi} \exp\left\{-\frac{1}{8\pi}\sum_{\ell\in\Gamma_1} 
    \Bigl[\d \varphi(\ell)\Bigr]^2\right\}}\,,
\end{equation}
\end{widetext}
where $n_{x,y}$ in the numerator is an arbitrary $\Z$-valued link field
that satisfies Eq.~\eqref{eq:dn=}.

To evaluate the remaining path integral over $\varphi$, we notice that
the link field $n_{x,y}$ in the numerator of
Eq.~\eqref{eq:S_intermediate} does not have to be $\Z$-valued. Since the
numerator of Eq.~\eqref{eq:S_intermediate} is invariant under the
transformation
\begin{equation}
    \varphi \to \varphi + \alpha \,,\qquad
    n_{x,y} \to n_{x,y} - \frac{\d\alpha}{2\pi}
\end{equation}
for any $\alpha:\Gamma_0\mapsto\R$, we can actually allow $n_{x,y}$ to
be any $\R$-valued link field that satisfies Eq.~\eqref{eq:dn=}. Then a
convenient choice is
\begin{equation}\label{eq:trick_magnetic}
    n_{x,y} \equiv - \delta\sigma_x + \delta\sigma_y
\end{equation}
for $\sigma_x:\Gamma_2\mapsto\R$ such that $\forall p\in\Gamma_2$,
\begin{equation}
    - \Delta \sigma_x(p) = \delta_{p,\star x}\,.
\end{equation}
This auxiliary plaquette field can be expressed in terms of the lattice
Green function we review in Appendix~\ref{app:LatticeGreenFunction},
\begin{equation}\label{eq:auxiliary_sigma}
    \sigma_x(p) = G(p-\star x)\,.
\end{equation}
Here $p-\star x$ has integer components and can thus be viewed as site
coordinates. Using the sum-by-parts identities~\eqref{eq:SumByParts},
the exactness $\d^2=\delta^2=0$, and the convention $G(0)=0$ of
Eq.~\eqref{eq:G_0_0_Convention}, we obtain 
\begin{widetext}
\begin{align}\label{eq:S_intermediate'}
    \frac{S_{\pm}(x-y)}{Z_{\pm}\overline{Z}_{\pm}} =
    \exp\Bigg\{\pi G(x-y)\,\pm & \,i\pi\sum_{z\in\Gamma_0} \d C_{x,y}(\star z) 
    \big[G\left(x - z\right) - G\left(y -z\right)\big]\Bigg\}\times\\ &
    \frac{\displaystyle \prod_{s\in\Gamma_0} \int_{\R}\frac{\d\varphi(s)}{2\pi} \exp\left\{-\frac{1}{8\pi}\sum_{\ell\in\Gamma_1} 
    \Bigl[\d \varphi(\ell)\Bigr]^2
    \pm\frac{i}{2}\left[\varphi(x) - \varphi(y)\right] \right\}}
    {\displaystyle \prod_{s\in\Gamma_0} \int_{\R}\frac{\d\varphi(s)}{2\pi} \exp\left\{-\frac{1}{8\pi}\sum_{\ell\in\Gamma_1} 
    \Bigl[\d \varphi(\ell)\Bigr]^2\right\}},\nonumber
\end{align}
\end{widetext}
We now introduce an auxiliary site field $\rho_x:\Gamma_0\mapsto\R$ such
that $\forall s\in\Gamma_0$,
\begin{equation}
    - \Delta \rho_x(s) = \delta_{s,x}\,.
\end{equation}
Again, it can be expressed in terms of the lattice Green function we
review in Appendix~\ref{app:LatticeGreenFunction},
\begin{equation}
    \rho_x(s) = G(s-x)\,.
\end{equation}
Using the sum-by-parts identities~\eqref{eq:SumByParts}, we can rewrite
\begin{align}\label{eq:trick_electric}
    \varphi(x) - \varphi(y) = \sum_{\ell} \d \varphi(\ell)\,\Big[\d\rho_x(\ell)-\d\rho_y(\ell)\Big]\,.
\end{align}
We can now complete the square in the numerator of
Eq.~\eqref{eq:S_intermediate'}:
introducing
\begin{equation}
    \bar{\varphi} \equiv \varphi \mp i2\pi\left(\rho_x-\rho_y\right)
\end{equation}
and using the convention $G(0)=0$, we obtain
\begin{widetext}
\begin{align}\label{eq:S_intermediate''}
    \frac{S_{\pm}(x-y)}{Z_{\pm}\overline{Z}_{\pm}} =
    \exp\Bigg\{2\pi G(x\!-\!y)\,\pm & \,i\pi\!\!\sum_{z\in\Gamma_0}\! \d C_{x,y}(\star z) 
    \big[G\left(x \!-\! z\right) - G\left(y \!-\! z\right)\!\big]\!\Bigg\}
    \frac{\displaystyle \prod_{s\in\Gamma_0} \int_{\R}\frac{\d\varphi(s)}{2\pi} \exp\left\{-\frac{1}{8\pi}\sum_{\ell\in\Gamma_1} 
    \Bigl[\d\bar{\varphi}(\ell)\Bigr]^2
    \right\}}
    {\displaystyle \prod_{s\in\Gamma_0} \int_{\R}\frac{\d\varphi(s)}{2\pi} \exp\left\{-\frac{1}{8\pi}\sum_{\ell\in\Gamma_1} 
    \Bigl[\d \varphi(\ell)\Bigr]^2\right\}},
\end{align}
\end{widetext}
By shifting $\varphi(s)$'s integral contour for all $s\in\Gamma_0$ based on the Gaussian function's complex analyticity and rapid convergence around real infinity, we see that the numerator and the denominator of Eq.~\eqref{eq:S_intermediate''} are identical to each other and thus cancel out.
We finally reach Eq.~\eqref{eq:2pt_with_G}.

\subsection{Asymptotic expansion}
\label{app:AsymptoticLatticeWeylCor}

In this subsection, we derive the large-$|x|$ asymptotic expansion of
the correlation function $S_{\pm}(x)$. Let us rewrite
Eq.~\eqref{eq:2pt_with_G} as
\begin{equation}
    S_{\pm}(x) = Z_{\pm}\overline{Z}_{\pm} \e^{2\pi G(x)} \e^{i\Phi_{\pm}(x)} \e^{i\Theta_{\pm}(x)}\,,
\end{equation}
where the phase functions 
\begin{subequations}\label{eq:phase}
\begin{gather}
    \e^{i\Phi_{\pm}(x)} \equiv \exp\left[\mp i\pi\!\sum_{\ell\in\Gamma_1}\!C_{x,0}(\ell) \delta \sigma_x(\ell)\right],\label{eq:phase_1} \\
    \e^{i\Theta_{\pm}(x)} \equiv \exp\left[\pm i\pi\!\sum_{\ell\in\Gamma_1}\!C_{x,0}(\ell)\delta \sigma_0(\ell)\right],\label{eq:phase_2}
\end{gather}
\end{subequations}
for the equivalence class of paths $C_{x,0}$ represented by
Fig.~\ref{fig:correct_branch} (and Fig.~\ref{fig:branch_consistency})
and the auxiliary plaquette fields $\sigma_x$ defined by
Eq.~\eqref{eq:auxiliary_sigma}. We can immediately find 
\begin{equation}\label{eq:WeylLatticeAsymMag}
    \e^{2\pi G(x)} \sim
    \frac{1}{2\sqrt{2}\e^{\gamma_E}} \left[\frac{1}{|x|} + \mathcal{O}\left(\frac{1}{|x|^3}\right)\right]
\end{equation}
according to $G(x)$'s asymptotic expansion~\eqref{eq:AsymptoticGreen}.
It remains to expand the phase functions.

\tikzmath{ \r = 4; } 
\tikzmath{ \t = 120; } 

\begin{figure}
\centering
\begin{tikzpicture}[scale=0.65]
    \coordinate (Start) at (0, 0); \coordinate (Middle) at (4.25, 4.25);
    \coordinate (End) at (-3.0, 5.25);

    \draw[gray!20, very thin, step=0.25] (-6,-2) grid (6,7);
    \draw[-,ultra thin] (-6,0)--(6,0); \draw[-,ultra thin]
    (0,-2)--(0,7);

    \fill[gray!40] (Start) rectangle +(0.25,0.25);
    \draw [very thick, DodgerBlue2]
    (0,0) -- (0.25,0) -- (0.25,0.25);
    \draw [very thick, DodgerBlue2]
    (0.25,0.25) -- (0.5,0.25) -- (0.5,0.5) -- (0.75,0.5) -- (0.75,0.75)
      -- (1.0,0.75) -- (1.0,1.0) -- (1.25,1.0) -- (1.25,1.25) --
      (1.5,1.25) -- (1.5,1.5) -- (1.75,1.5) -- (1.75,1.75) -- (2.0,1.75)
      -- (2.0,2.0) -- (2.25,2.0) -- (2.25,2.25) -- (2.5,2.25) --
      (2.5,2.5) -- (2.75,2.5) -- (2.75,2.75) -- (3.0,2.75) -- (3.0,3.0)
      -- (3.25,3.0) -- (3.25,3.25) -- (3.5,3.25) -- (3.5,3.5) --
      (3.75,3.5) -- (3.75,3.75) -- (4.0,3.75) -- (4.0,4.0) -- (4.25,4.0)
      -- (4.25,4.25);
    \draw [very thick, DodgerBlue2]
(4.25,4.25) -- (4.0,4.25) -- (4.0,4.5) -- (3.75,4.5) -- (3.75,4.75) --
  (3.5,4.75) -- (3.5,5.0) -- (3.25,5.0) -- (3.25,5.25) -- (3.0,5.25) --
  (2.75,5.25) -- (2.75,5.5) -- (2.5,5.5) -- (2.25,5.5) -- (2.25,5.75) --
  (2.0,5.75) -- (1.75,5.75) -- (1.5,5.75) -- (1.25,5.75) -- (1.0,5.75)
  -- (1.0,6.0) -- (0.75,6.0) -- (0.5,6.0) -- (0.25,6.0) -- (0.0,6.0) --
  (-0.25,6.0) -- (-0.5,6.0) -- (-0.75,6.0) -- (-1.0,6.0) -- (-1.0,5.75)
  -- (-1.25,5.75) -- (-1.5,5.75) -- (-1.75,5.75) -- (-2.0,5.75) --
  (-2.0,5.5) -- (-2.25,5.5) -- (-2.5,5.5) -- (-2.5,5.25) -- (-2.75,5.25)
  -- (-3.0,5.25);
    \fill[thick] (Start) circle (3pt); \fill[thick] (Middle) circle
    (3pt); \fill[thick] (End) circle (3pt); \node[below left] at (Start)
    {\large $0$}; \node[below right] at (Middle) {\large $x'$};
    \node[below left] at (End) {\large $x$};

    \node[below right, DodgerBlue2, fill=white, inner sep=1pt] at (2.75,
    2.25) {\large $C_{x'\!,0}$}; \node[above, DodgerBlue2, fill=white,
    inner sep=1pt] at (2, 4.5) {\large $C_{x,x'}$};
   
\end{tikzpicture}
\caption{The lattice path decomposition of
Eq.~\eqref{eq:path_decompose_1}. To select the right branch for the
path, it is important that the curve routes counterclockwise around the
shaded plaquette $\star0$, which corresponds to the $\sigma_0$ term in
Eq.~\eqref{eq:phase_2}.}
\label{fig:NegativeDyonWithDeformedLine}
\end{figure}
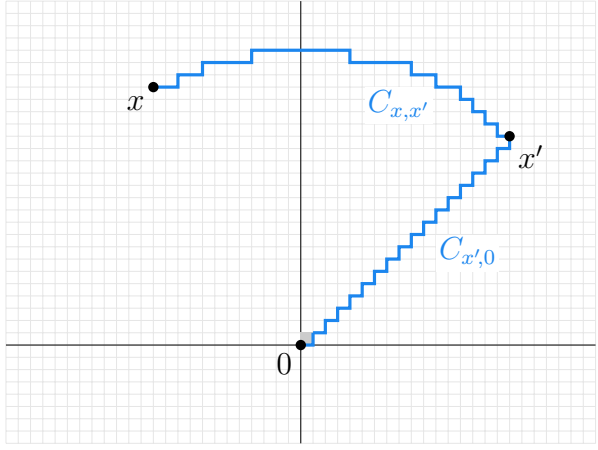

To evaluate the phase function $\e^{i\Theta_{\pm}(x)}$ for large $|x|$,
we can decompose the lattice path $C_{x,0}$ into two pieces,
\begin{equation}\label{eq:path_decompose_1}
    C_{x,0} = C_{x'\!,0} + C_{x,x'}\,,
\end{equation}
where
\begin{equation}
    x'\equiv\left(\left\lfloor\frac{|x|}{\sqrt{2}}\right\rfloor, \left\lfloor\frac{|x|}{\sqrt{2}}\right\rfloor\right).
\end{equation}
As illustrated by Fig.~\ref{fig:NegativeDyonWithDeformedLine}, the first
piece $C_{x',0}$ is a ``staircase'' path from the origin to $x'$, while
the second piece $C_{x,x'}$ is a lattice ``arc'' centered at the origin
stretching counterclockwise from $x'$ to $x$. 

Using the Poisson equation~\eqref{eq:PoissonEquation} and the lattice
$D_8$ symmetry, we can find that the $C_{x',0}$ contribution cancels out
on each ``stair'', except for the ``initial stair:''
\begin{equation}
\begin{gathered}
    \sum_{\ell\in\Gamma_1}\!C_{x'\!,0}(\ell)\delta \sigma_0(\ell) = \delta\sigma_0(\tfrac{1}{2},0) + \delta\sigma_0(1,\tfrac{1}{2}) \\
    = G(0,-1) + G(0,1) - 2G(0,0) = -\frac{1}{2}\,.
\end{gathered}
\end{equation}
Using $G(x)$'s asymptotic expansion~\eqref{eq:AsymptoticGreen}, we can
evaluate the asymptotic expansion of the $C_{x,x'}$ contribution. For
$x=(r\cos\theta,r\sin\theta)$, we find 
\begin{equation}
\begin{gathered}
    \sum_{\ell\in\Gamma_1}\!C_{x,x'}(\ell)\delta \sigma_0(\ell) \sim \int_{R_r\left(\tfrac{\pi}{4},\theta\right)} \boldsymbol{\nabla} G \times \d \boldsymbol{x} \\
    \sim -\frac{\theta}{2\pi} + \frac{1}{8} + \calO\left(\frac{1}{r^2}\right),
\end{gathered}
\end{equation}
where $R_r(\tfrac{\pi}{4},\theta)$ is a radius-$r$ arc with angles
counterclockwise from $\tfrac{\pi}{4}$ to $\theta$. Combining the two
contributions, we thus find
\begin{equation}\label{eq:WeylLatticeAsymArg_1}
    \e^{i\Theta_{\pm}(x)} \sim \e^{\mp i\left(\frac{\theta}{2} +\frac{3\pi}{8}\right)}\left[1+\calO\left(\frac{1}{r^2}\right)\right]
\end{equation}
for $x=(r\cos\theta,r\sin\theta)$.

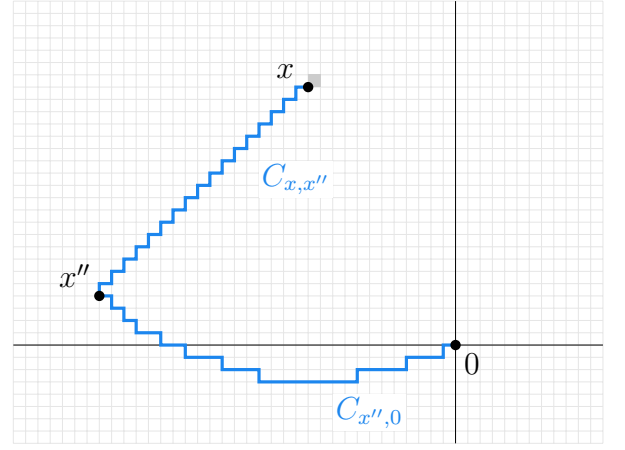
\begin{figure}
\centering
\begin{tikzpicture}[scale=0.65]
    \coordinate (Start) at (0, 0); \coordinate (Middle) at (-7.25, 1.0);
    \coordinate (End) at (-3.0, 5.25);

    \draw[gray!20, very thin, step=0.25] (-9,-2) grid (3,7);
    \draw[-,ultra thin] (-9,0)--(3,0); \draw[-,ultra thin]
    (0,-2)--(0,7);

    \fill[gray!40] (End) rectangle +(0.25,0.25);
    \draw [very thick, DodgerBlue2]
(0,0) -- (-0.25,0) -- (-0.25,-0.25) -- (-0.5,-0.25) -- (-0.75,-0.25) --
  (-1.0,-0.25) -- (-1.0,-0.5) -- (-1.25,-0.5) -- (-1.5,-0.5) --
  (-1.75,-0.5) -- (-2.0,-0.5) -- (-2.0,-0.75) -- (-2.25,-0.75) --
  (-2.5,-0.75) -- (-2.75,-0.75) -- (-3.0,-0.75) -- (-3.25,-0.75) --
  (-3.5,-0.75) -- (-3.75,-0.75) -- (-4.0,-0.75) -- (-4.0,-0.5) --
  (-4.25,-0.5) -- (-4.5,-0.5) -- (-4.75,-0.5) -- (-4.75,-0.25) --
  (-5.0,-0.25) -- (-5.25,-0.25) -- (-5.5,-0.25) -- (-5.5,0.0) --
  (-5.75,0.0) -- (-6.0,0.0) -- (-6.0,0.25) -- (-6.25,0.25) --
  (-6.5,0.25) -- (-6.5,0.5) -- (-6.75,0.5) -- (-6.75,0.75) --
  (-7.0,0.75) -- (-7.0,1.0) -- (-7.25,1.0);
    \draw [very thick, DodgerBlue2]
    (-7.25,1.0) -- (-7.25,1.25) -- (-7.0,1.25) -- (-7.0,1.5) --
      (-6.75,1.5) -- (-6.75,1.75) -- (-6.5,1.75) -- (-6.5,2.0) --
      (-6.25,2.0) -- (-6.25,2.25) -- (-6.0,2.25) -- (-6.0,2.5) --
      (-5.75,2.5) -- (-5.75,2.75) -- (-5.5,2.75) -- (-5.5,3.0) --
      (-5.25,3.0) -- (-5.25,3.25) -- (-5.0,3.25) -- (-5.0,3.5) --
      (-4.75,3.5) -- (-4.75,3.75) -- (-4.5,3.75) -- (-4.5,4.0) --
      (-4.25,4.0) -- (-4.25,4.25) -- (-4.0,4.25) -- (-4.0,4.5) --
      (-3.75,4.5) -- (-3.75,4.75) -- (-3.5,4.75) -- (-3.5,5.0) --
      (-3.25,5.0) -- (-3.25,5.25) -- (-3.0,5.25);
    \fill[thick] (Start) circle (3pt); \fill[thick] (Middle) circle
    (3pt); \fill[thick] (End) circle (3pt); \node[below right] at
    (Start) {\large $0$}; \node[above left] at (Middle) {\large $x''$};
    \node[above left] at (End) {\large $x\,$};

    \node[below right, DodgerBlue2, fill=white, inner sep=1pt] at (-2.5,
    -1) {\large $C_{x'',0}$}; \node[above, DodgerBlue2, fill=white,
    inner sep=1pt] at (-3.25, 3) {\large $C_{x,x''}$};
\end{tikzpicture}
\caption{The lattice path decomposition of
Eq.~\eqref{eq:path_decompose_2}. The shaded plaquette at $\star x$
corresponds to the term $\sigma_x$ in Eq.~\eqref{eq:phase_1}.}
\label{fig:PositiveDyonWithDeformedLine}
\end{figure}

We can evaluate the asymptotic expansion of $\e^{i\Phi_{\pm}(x)}$ in a
similar way. Let us decompose the lattice path $C_{x,0}$ into two
pieces,
\begin{equation}\label{eq:path_decompose_2}
    C_{x,0} = C_{x'\!'\!,0} + C_{x,x'\!'}\,,
\end{equation}
where
\begin{equation}
    x''\equiv x-\left(\left\lfloor\frac{|x|}{\sqrt{2}}\right\rfloor, \left\lfloor\frac{|x|}{\sqrt{2}}\right\rfloor\right).
\end{equation}
As illustrated by Fig.~\ref{fig:PositiveDyonWithDeformedLine}, the first
piece $C_{x'\!',0}$ is a lattice ``arc'' centered at $x$ stretching
clockwise from the origin to $x'\!'$, while the second piece
$C_{x,x'\!'}$ is a ``staircase'' path from $x'\!'$ to $x$. Using the
Poisson equation~\eqref{eq:PoissonEquation} and the lattice $D_8$
symmetry, we can find that the $C_{x,x'\!'}$ contribution completely
cancels out on each ``stair,'' leading to
\begin{equation}
\begin{gathered}
    \sum_{\ell\in\Gamma_1}\!C_{x,x'\!'}(\ell)\delta \sigma_x(\ell) = 0.
\end{gathered}
\end{equation}
Using $G(x)$'s asymptotic expansion~\eqref{eq:AsymptoticGreen}, we can
find for $x=(r\cos\theta,r\sin\theta)$, 
\begin{equation}
\begin{gathered}
    \sum_{\ell\in\Gamma_1}\!C_{x'\!'\!,0}(\ell) \delta\sigma_x(\ell) \sim \int_{R_r\left(\theta+\pi,\tfrac{5\pi}{4}\right)} \boldsymbol{\nabla} G \times \d \boldsymbol{x} \\
    \sim \frac{\theta}{2\pi} - \frac{1}{8} + \calO\left(\frac{1}{r^2}\right),
\end{gathered}
\end{equation}
where $R_r(\theta+\pi,\tfrac{5\pi}{4})$ is a radius-$r$ arc with angles
clockwise from $\theta+\pi$ to $\tfrac{5\pi}{4}$. Combining the two
contributions, we thus find
\begin{equation}\label{eq:WeylLatticeAsymArg_2}
    \e^{i\Phi_{\pm}(x)} \sim \e^{\mp i\left(\frac{\theta}{2} - \frac{\pi}{8}\right)}\left[1+\calO\left(\frac{1}{r^2}\right)\right]
\end{equation}
for $x=(r\cos\theta,r\sin\theta)$.

Finally, combining
Eqs.~\eqref{eq:WeylLatticeAsymMag},~\eqref{eq:WeylLatticeAsymArg_1},
and~\eqref{eq:WeylLatticeAsymArg_2}, we obtain the asymptotic expansion
of $S_{\pm}(x)$ as in Eq.~\eqref{eq:correlator_asymptotic}. 

\subsection{Momentum-space discontinuity}
\label{app:analytic_structure}

In this subsection, we discuss the singularities of the momentum-space
correlation function $\widetilde{S}_+(p)$. Since the leading
$\calO(|x|^{-1})$ term in the asymptotic
expansion~\eqref{eq:correlator_asymptotic} of $S_+(x)$ decays more
slowly than $\calO(|x|^{-2})$, the Fourier transform
$\widetilde{S}_+(p)$ will contain singularities.

We first show that $\widetilde{S}_+(p)$ is continuous on $T^2$ except
for $p=(0,0)$. Let us construct two auxiliary site fields
\begin{subequations}\label{eq:F&G}
\begin{gather}
    \begin{split}
        F(x_1,x_2) \equiv &\: S_+(x_1\!+\!1,x_2) - S_+(x_1,x_2) \\
        &\: +i S_+(x_1,x_2\!+\!1) -i S_+(x_1,x_2),
    \end{split}\\
    \begin{split}
        H(x_1,x_2) \equiv &\: S_+(x_1,x_2) - S_+(x_1\!-\!1,x_2) \\
        &\: +i S_+(x_1,x_2) -i S_+(x_1,x_2\!-\!1),
    \end{split}
\end{gather}
\end{subequations}
Using the asymptotic expansion~\eqref{eq:correlator_asymptotic}, by
explicit computation we can find that
\begin{equation}
    F(x) \sim \calO\left(\frac{1}{|x|^3}\right),\qquad
    H(x) \sim \calO\left(\frac{1}{|x|^3}\right).
\end{equation}
Since they decay faster than $|x|^{-2}$, their Fourier transforms
converge absolutely and thus uniformly. This further implies that their
Fourier transforms $\widetilde{F}(p)$ and $\widetilde{H}(p)$ are
continuous on $T^2$. By the definition~\eqref{eq:F&G} we have
\begin{subequations}
\begin{gather}
    \widetilde{F}(p) = \left[(\e^{ip_1} \!-\! 1)+i(\e^{ip_2} \!-\! 1)\right] \widetilde{S}_+(p),\\
    \widetilde{H}(p) = \left[(1 \!-\! \e^{-ip_1})+i(1 \!-\! \e^{-ip_2})\right] \widetilde{S}_+(p).
\end{gather}
\end{subequations}
The functions in front of $\widetilde{S}_+(p)$ above are both smooth on
$T^2$. The former function $(\e^{ip_1} \!-\! 1)+i(\e^{ip_2} \!-\! 1)$
has two zeros located at
\begin{equation}\label{eq:zero_F}
    (0,0)\quad\text{and}\quad\left(\pi/2,-\pi/2\right).
\end{equation}
The latter function $(1 \!-\! \e^{-ip_1})+i(1 \!-\! \e^{-ip_2})$ has two
zeros located at
\begin{equation}\label{eq:zero_G}
    (0,0)\quad\text{and}\quad\left(-\pi/2,\pi/2\right),
\end{equation}
Since we have proved that $\widetilde{F}(p)$ and $\widetilde{H}(p)$ are
continuous, $\widetilde{S}_+(p)$ must also be continuous except at the
common zeros of Eqs.~\eqref{eq:zero_F} and~\eqref{eq:zero_G}. Namely,
$\widetilde{S}_+(p)$ is continuous except at $p=(0,0)$.

We now clarify the form of the discontinuity by finding the small-$|p|$
asymptotic expansion of $\widetilde{S}_+(p)$. Let us express the
position-space $S_+(x)$ by
\begin{equation}
    S_+(x) = R_+(x) + \left\{\begin{aligned}
        &\frac{1}{x_1\!+\!ix_2}, &&x\neq0\\
        &0, &&x=0
    \end{aligned}\right..
\end{equation}
According to Eq.~\eqref{eq:correlator_asymptotic}, we have
\begin{equation}
    R_+(x) \sim \calO\left(\frac{1}{|x|^3}\right),
\end{equation}
and therefore, $\widetilde{R}_+(p)$ is continuous on $T^2$. Accordingly,
in momentum space, we have
\begin{equation}
    \widetilde{S}_+(p) = \widetilde{R}_+(p) - 2\pi i \left[\zeta_{\Lambda}(p_1\!+\!ip_2) - \frac{p_1\!-\!ip_2}{4\pi}\right],
\end{equation}
where $\Lambda\subset\C$ is the lattice 
\begin{equation}
    \Lambda \equiv 2\pi(\Z + i\Z) \subset\C
\end{equation}
and $\zeta_{\Lambda}(z)$ is the Weierstrass zeta function for $\Lambda$,
\begin{equation}
    \zeta_{\Lambda}(z) = \frac{1}{z} + \!\!\!\sum_{w\in\Lambda-\{0\}}\!\left(\frac{1}{z-w} + \frac{1}{w} + \frac{z}{w^2}\right)\,.
\end{equation}
Clearly, $\zeta_{\Lambda}(z)$ is a meromorphic function on $\C$ with
simple poles on $z\in\Lambda$ with unit residues. We thus obtain the
small-$|p|$ asymptotic expansion~\eqref{eq:pole}.


\bibliography{main}

\end{document}